\documentclass[a4paper,fleqn]{cas-dc}

\usepackage[authoryear]{natbib}
\usepackage{amsmath} 
\usepackage{amsfonts} 
\usepackage{float} 
\usepackage{algorithm}
\usepackage{algpseudocode}
\usepackage{hyperref}

\def\tsc#1{\csdef{#1}{\textsc{\lowercase{#1}}\xspace}}
\tsc{WGM}
\tsc{QE}
\tsc{EP}
\tsc{PMS}
\tsc{BEC}
\tsc{DE}
\begin{document}
\let\WriteBookmarks\relax
\def\floatpagepagefraction{1}
\def\textpagefraction{.001}
\shorttitle{}
\shortauthors{Wenjie Wang et~al.}

\title [mode = title]{Adaptive Deep Koopman Operator for Vehicle Dynamics Modeling: A Physics-Informed and Tire-Force-Driven Approach}                      

\author[1,2]{Wenjie Wang}[style=chinese]
\credit{Methodology, Software, Validation, Formal analysis, Investigation, Visualization, Writing-original draft, Writing-review \& editing}

\address[1]{College of Mechanical and Vehicle Engineering, Chongqing University, Chongqing, 400044, China}
\address[2]{Department of Automotive Engineering, Hanyang University, Seoul, 04763, South Korea}
\address[3]{School of Vehicle Engineering, Chongqing University of Technology, Chongqing, 401135, China}

\author[1]{Hao Chen}[style=chinese]
\credit{Conceptualization, Writing-review \& editing}

\author[3]{Ran Shu}[style=chinese]
\credit{Conceptualization, Writing-review \& editing}

\author[2]{Solyeon Kwon}[style=chinese]
\credit{Data Curation, Validation}

\author[2]{Kyoungseok Han}[style=chinese]
\credit{Resources, Writing-review \& editing}

\author[1]{Hongyu Shu}[style=chinese,orcid=0000-0002-5607-5103]
\cormark[1]
\ead{shycqu@cqu.edu.cn}
\cortext[cor1]{Corresponding author}	
\credit{Funding acquisition, Conceptualization, Supervision}

\begin{abstract}
Accurate and adaptive modeling of vehicle dynamics is paramount for the safety of autonomous driving systems, particularly under extreme maneuvers and time-varying parameters. While Deep Koopman operator theory offers a promising global linearization framework, its online application faces a theoretical bottleneck: the high-dimensional lifted state space inherently induces a rank-deficient problem, rendering traditional recursive least squares based updates numerically unstable. To address this, we propose a novel tire-force-driven modeling framework with guaranteed online stability. First, an offline Deep Koopman model is constructed by embedding 7DOF dynamic equilibrium constraints into the learning objective, ensuring the structural fidelity and physical interpretability of the lifted manifold. Second, we theoretically reformulate the operator update in the rank-deficient space as a minimum-norm solution problem. A Physics-Informed Variable Step-Size Normalized Least Mean Squares (PI-VSS-NLMS) algorithm is proposed, which leverages the projection property of NLMS to act as a stable pseudo-inverse solver while incorporating an anchoring mechanism to suppress parameter drift. Extensive simulations on CarSim and Hardware-in-the-Loop validation on dSPACE MicroAutobox III confirm the superiority of the proposed algorithm. It achieves robust prediction accuracy under unseen excitations while guaranteeing real-time feasibility with an average execution time of 0.421 ms, thus bridging the gap between theoretical models and practical deployment.
\end{abstract}

\maketitle

\section{Introduction}
\label{sec:1}
Precise state prediction is paramount for autonomous vehicle stability, particularly given the strong nonlinearities and time-varying parameters near handling limits. This requires a modeling framework that balances global nonlinear accuracy with real-time adaptability.

Existing modeling approaches generally fall into two categories. Physics-based models (e.g., seven-degree-of-freedom (7DOF)  \citep{liu2026stability}, 14DOF \citep{yu2018optimal}, 17DOF \citep{yang2023new}) offer interpretability, but are sensitive to parameter uncertainties and unmodeled dynamics. Moreover, maintaining their accuracy via calibration or adaptation imposes high computational costs and is prone to inherent uncertainties. In contrast, data-driven models bypass the need for explicit parameter identification and structural modeling. For instance, in the context of autonomous driving control, \citet{spielberg2019} demonstrated that a neural network vehicle model could outperform physics-based ones, enabling expert-level trajectory tracking without explicit friction estimation. Meanwhile, for applications demanding high-fidelity dynamics modeling, \citet{pan2021data} proposed a deep neural network trained on multibody simulation data for accurate longitudinal dynamics prediction, while \citet{hermansdorfer2021end} developed an end-to-end architecture combining neural networks with kinematics to capture nonlinear characteristics. Despite their high accuracy, such models intrinsically lack interpretability due to the absence of physical constraints, and their performance is strictly confined by training data, leading to poor generalization in unseen scenarios.

The Koopman operator theory \citep{koopman1931hamiltonian} offers a pathway to bridge the gap between data-driven accuracy and control-oriented linearity by lifting finite-dimensional nonlinear dynamics into an infinite-dimensional function space where the evolution becomes globally linear. For practical engineering applications, dynamic mode decomposition (DMD) \citep{rowley2017model} and extended DMD \citep{williams2016extending}  were developed to project the infinite-dimensional dynamics onto a finite subspace spanned by a set of observable functions. However, the performance of these methods is often suboptimal for highly coupled, nonlinear vehicle dynamics due to their reliance on manually designed basis functions.

To overcome the limitations of manual feature engineering, Deep Koopman architectures have emerged, employing a deep neural network(DNN) to automatically learn optimal finite-dimensional lifting functions \citep{yeung2019learning}. By unifying the learning of coordinate embeddings and evolution operators into an end-to-end framework \citep{shi2022deep}, these methods effectively global-linearize coupled vehicle dynamics, thereby facilitating efficient model predictive control (MPC) implementations \citep{xiao2022deep}. This paradigm has been successfully applied on various platforms, ranging from autonomous racing \citep{wang2021deep} to heavy-duty tracked vehicles \citep{zuo2025model}, demonstrating significant gains in both computational efficiency and trajectory tracking fidelity. Furthermore, to strictly guarantee operational safety, \citet{chen2024deep} integrated Deep Koopman operators with control barrier functions to ensure vehicle lateral stability through a safety command governor. To address the absence of intrinsic physical causality in purely data-driven models, the concept of physics-informed neural networks (PINNs) \citep{cai2021physics} has been introduced to regularize the learning process, where \citet{zhang2026physics} embedded acceleration-based physical loss by penalizing discrepancies between predicted and measured accelerations. Additionally, to account for time-varying parameters such as vehicle mass or road friction coefficients, online adaptation mechanisms are often employed, typically utilizing adaptive filters such as recursive least squares (RLS) \citep{abraham2019active, sayed2024recursive}, forgetting factor recursive least squares (FFRLS) \citep{calderon2021koopman, junker2023adaptive} or sliding window least squares (SWLS) \citep{zhang2026physics} to update the Koopman matrices in real-time.

Despite these advances, two fundamental bottlenecks hinder the robust deployment of Deep Koopman models in autonomous driving: first, the misalignment between physical constraints and dynamic causality, where reliance on kinematic effects (e.g., acceleration) rather than the causal tire-road interaction forces obscures the generation mechanism and degrades fidelity under force saturation; second, the theoretical incompatibility between high-dimensional lifting and traditional online adaptation, where applying methods like RLS face inherent rank deficiency—as the lifted state dimension vastly exceeds the intrinsic dynamics dimension—thus failing to balance tracking speed and numerical stability upon matrix rank collapse.

To bridge these gaps, this paper proposes a tire-force-driven Deep Koopman framework designed for robust and stable online adaptation. Departing from kinematic approximations, the core premise is to establish force-level causality within the Koopman lifting process, thereby ensuring physical interpretability and structural fidelity of the learned latent space. The framework is practically oriented: while it aligns with next-generation intelligent chassis technologies (e.g., advanced wheel force sensors\citep{xu2025intelligent, 10918823}), it remains fully compatible with standard tire-force observers \citep{ray1997nonlinear, baffet2009estimation, viehweger2021vehicle}, enabling deployment on production vehicles. The main contributions of this paper are summarized as follows:
\begin{enumerate}[(1)] 
\item We propose a novel Deep Koopman model that explicitly embeds 7DOF dynamic equilibrium constraints into the learning objective. This force-driven design shifts the modeling paradigm from kinematic fitting to dynamic causation, effectively eliminating non-physical artifacts and significantly improving generalization in nonlinear regimes. 
\item We develop a Physics-Informed Variable Step-Size Normalized Least Mean Squares (PI-VSS-NLMS) algorithm to address the online instability caused by the inherent rank deficiency of the lifted state space. By reformulating the update as a minimum-norm solution problem, this method circumvents the singularity accumulation of RLS-based methods, guaranteeing bounded stability and robust noise suppression during high-dynamic maneuvers. 
\end{enumerate}

The remainder of this paper is organized as follows: Section~\ref{sec:2} establishes theoretical preliminaries. Section~\ref{sec:3} details the offline physics-informed Deep Koopman model and the online PI-VSS-NLMS adaptation algorithm. Section~\ref{sec:4} presents the experimental validation and discussion. Section~\ref{sec:5} concludes the paper.

\section{Preliminaries}
\label{sec:2}
\subsection{Vehicle Dynamics Model}
\label{sec:2.1}
\begin{figure}
	\centering
	\includegraphics[width=\linewidth]{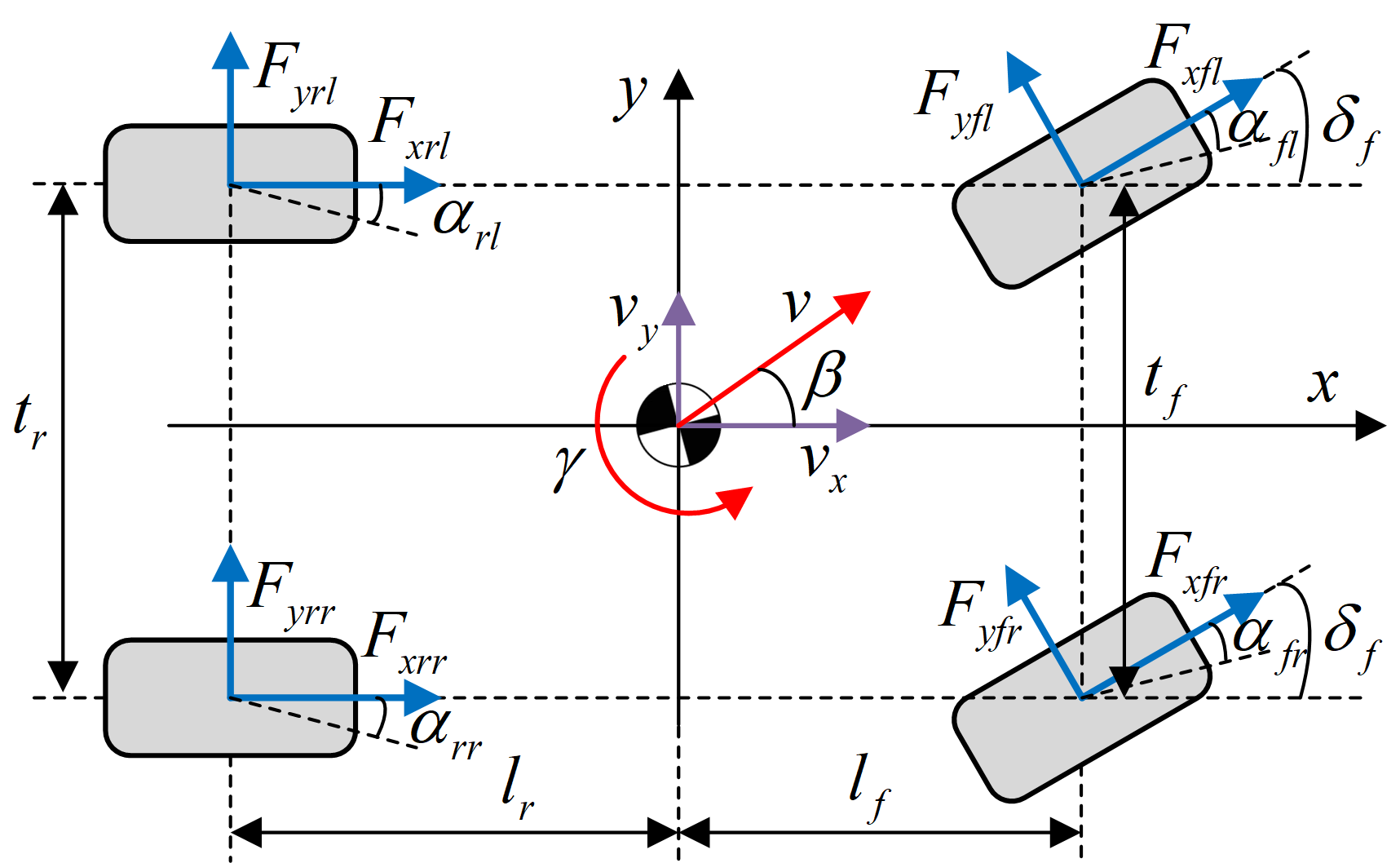}
	\caption{Planar vehicle dynamics model.}
    \label{fig1}
\end{figure}
To ensure physical consistency, the data-driven framework is anchored by a 7DOF vehicle model. As depicted in Fig.~\ref{fig1}, the chassis is modeled as a rigid body with planar motion defined by longitudinal velocity, lateral velocity, and yaw rate, while the remaining 4DOF account for individual wheel rotations. The governing Newton-Euler equations are formulated as follows:
\begin{align}
    \label{1}
    \hspace{-20pt}
	\left\{
	\begin{aligned}
	\dot{v}_x & = \frac{1}{m} \big[ (F_{xfl} + F_{xfr}) \cos \delta_f - (F_{yfl} + F_{yfr}) \sin \delta_f \\
	&\quad + F_{xrl} + F_{xrr} \big]+ v_y \gamma  \\
	\dot{v}_y & = \frac{1}{m} \big[ (F_{xfl} + F_{xfr}) \sin \delta_f + (F_{yfl} + F_{yfr}) \cos \delta_f \\
	&\quad + F_{yrl} + F_{yrr} \big]- v_x \gamma  \\
	\dot{\gamma} & = \frac{t_f}{2 I_z} \big[ (F_{xfr} - F_{xfl}) \cos \delta_f + (F_{yfl} - F_{yfr}) \sin \delta_f \big] \\
	&\quad + \frac{l_f}{I_z} \big[ (F_{xfl} + F_{xfr}) \sin \delta_f + (F_{yfl} + F_{yfr}) \cos \delta_f \big] \\
	&\quad + \frac{t_r}{2 I_z} ( F_{xrr} - F_{xrl} ) - \frac{l_r}{I_z} ( F_{yrl} + F_{yrr} )
	\end{aligned}
	\right.
\end{align}

\noindent 
where $m$, $v_x$, $v_y$, $\gamma$, $l_f$, $l_r$, $I_z$ are the vehicle mass, the longitudinal/lateral velocity at the center of gravity of the vehicle, the yaw rate, the distance from the front/rear axle to the center of gravity and the yaw inertia, respectively. $\delta_f = \delta_{sw}/i_{sw}$ represents the steering angle of the front wheel, with $\delta_{sw}$ denoting the steering wheel angle and $i_{sw}$ representing the steering ratio. $t_f$, $t_r$ are the track width of the front/rear axle, respectively. $F_{xij}$ and $F_{yij}$ (ij = fl, fr, rl, rr, representing front-left, front-right, rear-left and rear-right, respectively) represent longitudinal forces and lateral forces of the tires, which can be calculated through tire formulas \citep{pacejka2005tire, dugoff1970analysis,deur20043d}.

The state and input vectors are defined as:
\begin{equation}
\label{2}
\begin{aligned}
\boldsymbol{x} &= \left[ v_x \quad v_y \quad \gamma \right]^\top \\
  \boldsymbol{u} &= \left[ T \quad \delta_{sw} \right]^\top
\end{aligned}
\end{equation}
\noindent 
where $T$ represents the total torque.

Based on Eq.~\eqref{2}, the nonlinear discrete-time form of Eq.~\eqref{1} can be expressed as Eq.~\eqref{3}.
\begin{equation}
\label{3}
    \begin{bmatrix}
        v_{x, k+1} \\
        v_{y, k+1} \\
        \gamma_{k+1}
    \end{bmatrix}
    = f \left(
    \begin{bmatrix}
        v_{x, k} \\
        v_{y, k} \\
        \gamma_{k}
    \end{bmatrix},
    \begin{bmatrix}
        T_k \\
        \delta_{sw, k}
    \end{bmatrix}
    \right)
\end{equation}

\subsection{Basic Koopman Operator Theory}
\label{sec:2.2}
Koopman operator theory provides a powerful framework for lifting the nonlinear dynamics of a system into a linear representation within an infinite-dimensional function space. The $n$-dimensional discrete nonlinear system can be described as Eq.~\eqref{4}.
\begin{equation}
\label{4}
\boldsymbol{x}_{k+1} = \boldsymbol f(\boldsymbol{x}_k, \boldsymbol{u}_k)
\end{equation}
\noindent 
where $k \in \mathbb{N}$ denotes the discrete time step, $x_k \in \mathbb{R}^n$ is the $n$-dimensional state vector at time $k$, $u_k \in \mathbb{R}^m$ is the $m$-dimensional control vector at time $k$, and $f:\mathbb{R}^{n+m} \to \mathbb{R}^n$ represents the nonlinear mapping that describes the state transition between consecutive time steps.

By lifting the dynamics into an infinite-dimensional space via the observation function $\boldsymbol g(\cdot)$, the Koopman operator $\mathcal{K}$ enables a globally linear evolution, as described in Eq.~\eqref{5}.
\begin{equation}
\label{5}
\mathcal{K} \cdot \boldsymbol g(\boldsymbol{x}_k, \boldsymbol{u}_k) = \boldsymbol g(\boldsymbol{x}_{k+1}, \boldsymbol{u}_{k+1}) = \boldsymbol g(\boldsymbol f(\boldsymbol{x}_k, \boldsymbol{u}_k), \boldsymbol{u}_{k+1})
\end{equation}
\noindent 
where $\boldsymbol g: \mathbb{R}^{n+m} \to \mathbb{R}^d$ is a smooth vector field of $\mathrm {C}^{\infty}$.

The observation function $\boldsymbol g(\cdot)$ can be further separated into two parts in terms of states and control input, as shown in Eq.~\eqref{6}.
\begin{equation}
\label{6}
\boldsymbol g(\boldsymbol{x}_k, \boldsymbol{u}_k) = \begin{bmatrix} \boldsymbol g_x(\boldsymbol{x}_k) \\ \boldsymbol{u}_k \end{bmatrix}
\end{equation}
\noindent 
where $\boldsymbol g_x: \mathbb{R}^{n} \to \mathbb{R}^{d-m}$.

However, the implementation of infinite-dimensional $\mathcal{K}$ is impractical, but it can be realized through a finite-dimensional approximation \citep{brunton2021modern}. Thus, Eq.~\eqref{5} is simplified as Eq.~\eqref{7}:
\begin{equation}
\label{7}
\begin{bmatrix}
  \boldsymbol g_x(\boldsymbol{x}_{k+1}) \\
  \boldsymbol{u}_{k+1}
\end{bmatrix}
=
\begin{bmatrix}
  \boldsymbol{K}_{xx} & \boldsymbol{K}_{xu} \\
  \boldsymbol{K}_{ux} & \boldsymbol{K}_{uu}
\end{bmatrix}
\begin{bmatrix}
  \boldsymbol g_x(\boldsymbol{x}_k) \\
  \boldsymbol{u}_k
\end{bmatrix}
\end{equation}
\noindent 
where $\boldsymbol{K} \in \mathbb{R}^{d \times d}$ is the finite-dimensional approximation of $\mathcal{K}$ in the lifted space.

Since the system control input $\boldsymbol{u}_k$ does not need to be predicted, Eq.~\eqref{7} can be rewritten as Eq.~\eqref{8}.
\begin{equation}
\label{8}
\boldsymbol{z}_{k+1} = \boldsymbol{A}\boldsymbol{z}_k + \boldsymbol{B}\boldsymbol{u}_k
\end{equation}
\noindent 
where $\boldsymbol{A} = \boldsymbol{K}_{xx} \in \mathbb{R}^{(d-m) \times (d-m)}$ and $\boldsymbol{B} = \boldsymbol{K}_{xu} \in \mathbb{R}^{(d-m) \times m}$ are the system and the input matrices in the lifted space, respectively.

The component $\boldsymbol{z}_k = \boldsymbol g_x$ of the observation function $\boldsymbol g(\cdot)$ is defined as Eq.~\eqref{9}.
\begin{equation}
\label{9}
\boldsymbol{z}_k = \boldsymbol g_x(\boldsymbol{x}_k) = \begin{bmatrix} \boldsymbol{x}_k \\ \boldsymbol{\Phi}(\boldsymbol{x}_k) \end{bmatrix}
\end{equation}
where $\boldsymbol{\Phi}(\boldsymbol{x}_k) = [\boldsymbol{\varphi}_{1,k}(\boldsymbol{x}_k), \boldsymbol{\varphi}_{2,k}(\boldsymbol{x}_k), \dots, \boldsymbol{\varphi}_{d-m-n,k}(\boldsymbol{x}_k)]^\top \in \mathbb{R}^{d-m-n}$ is the basis in the lifted space.

Since $\boldsymbol{z}_k$ explicitly contains $\boldsymbol{x}_k$, the original states can be recovered via a linear projection:
\begin{equation}
\label{10}
\boldsymbol{x}_k = \boldsymbol{C}\boldsymbol{z}_k
\end{equation}
with the selection matrix $\boldsymbol{C} = [ \boldsymbol{I}_n \quad \boldsymbol{0} ]$.

Furthermore, the $l$-snapshot collections from the lifted space and inputs are as shown in Eq.~\eqref{11}
\begin{equation}
\label{11}
\begin{aligned}
\boldsymbol{Z}_X &= [\boldsymbol{z}_1, \boldsymbol{z}_2, \dots, \boldsymbol{z}_l]_{(d-m) \times l} \\
\boldsymbol{U} &= [\boldsymbol{u}_1, \boldsymbol{u}_2, \dots, \boldsymbol{u}_l]_{m \times l} \\
\boldsymbol{Z}_Y &= [\boldsymbol{z}_2, \boldsymbol{z}_3, \dots, \boldsymbol{z}_{l+1}]_{(d-m) \times l}
\end{aligned}
\end{equation}

The Koopman matrices are then determined by solving the following least-squares problem:
\begin{equation}
\label{12}
\begin{bmatrix} \boldsymbol{A} & \boldsymbol{B} \end{bmatrix} = \arg \min_{\boldsymbol{A}, \boldsymbol{B}} \left\| \boldsymbol{Z}_Y - \boldsymbol{A}\boldsymbol{Z}_X - \boldsymbol{B}\boldsymbol{U} \right\|_2^2 
\end{equation}

The analytical solution is given by the following.
\begin{equation}
\label{13}
\begin{bmatrix} \boldsymbol{A} & \boldsymbol{B} \end{bmatrix} = \boldsymbol{Z}_Y \begin{bmatrix} \boldsymbol{Z}_X \\ \boldsymbol{U} \end{bmatrix}^\top \left( \begin{bmatrix} \boldsymbol{Z}_X \\ \boldsymbol{U} \end{bmatrix} \begin{bmatrix} \boldsymbol{Z}_X \\ \boldsymbol{U} \end{bmatrix}^\top \right)^{\dagger}
\end{equation}
\noindent 
where $\dagger$ denotes the pseudo-inverse.

\subsection{Deep Koopman Operator Theory}
\label{sec:2.3}
Although the least-squares solution in Eq.~\eqref{13} is computationally efficient, the representational power of the lifted system is strictly dependent on the selection of the basis function $\boldsymbol{\Phi}$. To avoid the subjectivity and limitations of manual feature engineering, this study employs a data-driven approach based on DNN to automatically construct the coordinate transformation.

Specifically, the Deep Koopman architecture is realized via an auto-encoder framework \citep{shi2022deep}, where a trainable encoder maps the original nonlinear states into a high-dimensional latent manifold governed by linear evolution. In this structure, the encoder and decoder, which approximate the lifting mapping $\boldsymbol{\Phi}$ and its inverse $\boldsymbol{\Phi}^{-1}$, are implemented as multi-layer perceptrons \citep{riedmiller2014multi}. The state propagation through the $j$-th hidden layer of these networks is formulated as:
\begin{equation}
\label{14}
\boldsymbol{y}_j = \boldsymbol\sigma_j(\boldsymbol{W}_j \boldsymbol{y}_{j-1} + \boldsymbol{b}_j) \quad (j = 1, \dots, J)
\end{equation}
\noindent 
where $\boldsymbol{W}_j \in \mathbb{R}^{n_j \times n_{j-1}}$ and $\boldsymbol{b}_j \in \mathbb{R}^{n_j}$ denote the weight matrix and bias vector of the $j$-th layer, while $\boldsymbol{\sigma}_j$ and $n_j$ represent the activation function and layer width, respectively.

Complementing the auto-encoder structure, the system matrices $\boldsymbol{A}$ and the input matrices$\boldsymbol{B}$ are integrated as trainable linear layers without activation functions or biases. During training, the encoder $\boldsymbol\Phi$, the decoder $\boldsymbol\Phi^{-1}$, the system matrix $\boldsymbol{A}$ and the input matrix $\boldsymbol{B}$ are updated synchronously to minimize a composite loss function, which typically consists of three components:

\begin{enumerate}[(1)]
\item \textbf{Linearity loss:} designed to ensure the accuracy of linear state evolution within the lifted high-dimensional space.
\begin{equation}
\label{15}
\mathcal{L}_{\text{Lin}} = \frac{1}{N_p}\sum_{i=1}^{p} \xi^{i-1}\left\|
\begin{bmatrix}
  \boldsymbol{x}_{i+1} \\
  \boldsymbol{\Phi}(\boldsymbol{x}_{i+1})
\end{bmatrix}
-
\begin{bmatrix}
  \boldsymbol{A} & \boldsymbol{B}
\end{bmatrix}
\begin{bmatrix}
  \begin{bmatrix}
   \boldsymbol{x}_i \\
  \boldsymbol{\Phi}(\boldsymbol{x}_i)
  \end{bmatrix}\\
  \boldsymbol{u}_i
\end{bmatrix}
\right\|_2^2
\end{equation}
\noindent 
where $p$ denotes the prediction horizon, and $\xi \in (0, 1]$ is the discount factor, and $N_p = \sum_{i=1}^p \xi^{i-1}$ is the normalization term.
\item \textbf{Reconstruction loss:} introduced to ensure the consistency of the original state information after reconstruction by the encoder and decoder.
\begin{equation}
\label{16}
\mathcal{L}_{\text{Recon}} = \frac{1}{N_p}\sum_{i=1}^{p} \xi^{i-1}\left\| \boldsymbol{x}_i - \boldsymbol{\Phi}^{-1}(\boldsymbol{\Phi}(\boldsymbol{x}_i)) \right\|_2^2 
\end{equation}
\item \textbf{Prediction loss:} defined to verify the prediction accuracy of the future states for the nonlinear system.
\begin{equation}
\label{17}
\mathcal{L}_{\text{Pred}} = \frac{1}{N_p}\sum_{i=1}^{p} \xi^{i-1}\left\| 
\boldsymbol{x}_{i+1} - \boldsymbol{C} 
\begin{bmatrix} 
\boldsymbol{A} & \boldsymbol{B} 
\end{bmatrix} 
\begin{bmatrix} 
  \begin{bmatrix}
   \boldsymbol{x}_i \\
   \boldsymbol{\Phi}(\boldsymbol{x}_i)
  \end{bmatrix}\\
\boldsymbol{u}_i \end{bmatrix} \right\|_2^2 
\end{equation}
\end{enumerate}
\noindent 
\textbf{Remark 1.} The joint optimization of Eqs.~\eqref{15}--\eqref{17} enables the DNN to autonomously learn an optimal coordinate transformation for finite-dimensional Koopman approximation.

\section{Methodology}
\label{sec:3}
\subsection{Offline Modeling: Physics-Informed Deep Koopman}
\begin{figure*}
	\centering
	\includegraphics[width=0.8\linewidth]{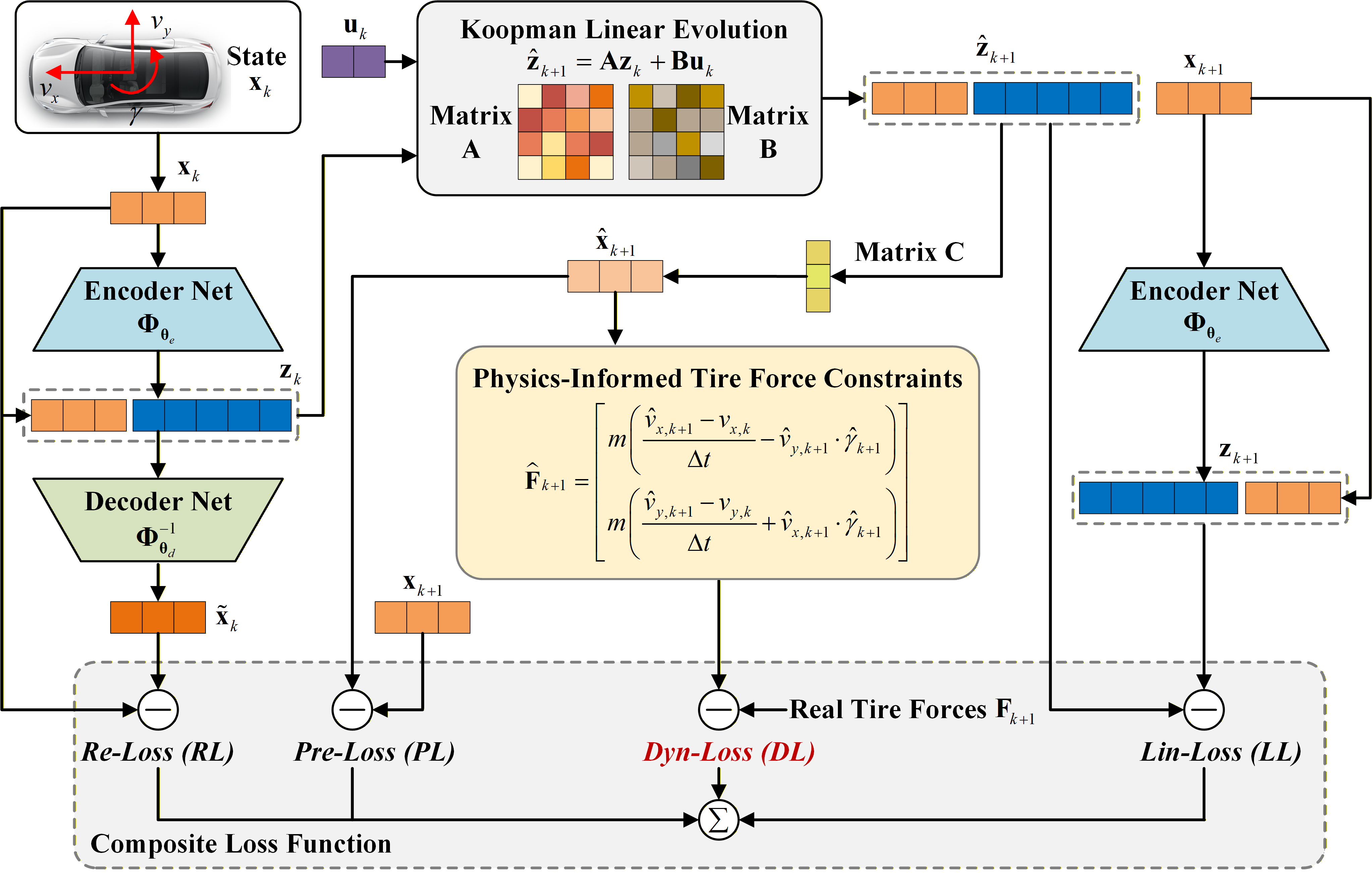}
	\caption{Architecture of the proposed Dynamic Loss-informed Deep Koopman (DLDK) framework.}
	\label{fig2}
\end{figure*}

This section delineates the proposed tire-force-driven offline modeling framework. As illustrated in Fig.~\ref{fig2}, this framework implements a tire-force-driven modeling philosophy by explicitly minimizing the force equilibrium residual. The core innovation lies in embedding vehicle kinetics equilibrium as a prior inductive bias into the deep learning optimization process, thereby ensuring that the learned dynamics strictly adhere to physical laws.

\subsubsection{Model Architecture}
Following the framework outlined in Section~\ref{sec:2.3}, a deep auto-encoder is employed to learn the Koopman observables. The encoder $\boldsymbol\Phi(\cdot; \boldsymbol{\theta}_e)$ functions as a nonlinear coordinate transformation, embedding the physical state $\boldsymbol{x}_k = [v_{x,k}, v_{y,k}, \gamma_k]^\top$ into a high-dimensional latent space $\mathbb{R}^{d-m}$. Conversely, the decoder $\boldsymbol\Phi^{-1}(\cdot; \boldsymbol{\theta}_d)$ performs the inverse mapping to ensure state reconstructability. Within this lifted space, the nonlinear evolution of vehicle states is approximated by the learnable matrices $\boldsymbol{A}$ and $\boldsymbol{B}$.

Consistently with the definitions in Eq.~\eqref{8}, the forward propagation is given by:
\begin{equation}
\label{18}
\hat{\boldsymbol{z}}_{k+1} = \boldsymbol{A}\boldsymbol{z}_k + \boldsymbol{B}\boldsymbol{u}_k, \quad \hat{\boldsymbol{x}}_{k+1} = \boldsymbol{C}\hat{\boldsymbol{z}}_{k+1}
\end{equation}

\subsubsection{Physics-Informed Loss Function}
\label{sec:loss_function}

To train network parameters $[\boldsymbol{\theta}_e, \boldsymbol{\theta}_d, \boldsymbol{A}, \boldsymbol{B}]$, we design a composite loss function that balances data-driven accuracy with physical consistency.

\begin{enumerate}[(1)]
\item \textbf{Canonical Koopman Losses:} A set of canonical Koopman losses were adopted to guarantee the mathematical validity of the operator within the lifted manifold, including linearity loss, reconstruction loss, and prediction loss. Collectively, these terms ensure that the latent space is bi-directionally mapped to the physical domain and that the state evolution adheres to linear transition laws.

\item \textbf{Physics-Informed Dynamic Loss:} The Physics-Informed Dynamic Loss ($\mathcal{L}_{\text{DL}}$), derived from the Eq.~\eqref{1}, serves as a structural inductive bias that enforces physical consistency within the data-driven framework. By penalizing the residuals between predicted inertial motion and measured tire forces, $\mathcal{L}_{\text{DL}}$ quantifies any violation of force equilibrium. Eq.~\eqref{19} ensures that the Koopman operator respects physical causality and remains robust against non-physical noise in the training data.

\begin{equation}
\label{19}
\mathcal{L}_{\text{DL}} = \frac{1}{N_p}\sum_{i=1}^{p} \xi^{i-1}\left\| \boldsymbol{F}_{k+i} - \hat{\boldsymbol{F}}_{k+i} \right\|_2^2  
\end{equation}
\noindent 
where $\hat{\boldsymbol{F}}_{k+i}$ represents inertial force vector at the $i$-th prediction step. By discretizing the differential equations in Eq.~\eqref{1}, $\hat{\boldsymbol{F}}_{k+i}$ is derived as:

\begin{equation}
\label{20}
\hat{\boldsymbol{F}}_{k+i} = 
\begin{bmatrix}
m \left( \frac{\hat{v}_{x,k+i} - \hat{v}_{x,k+i-1}}{\Delta t} - \hat{v}_{y,k+i} \cdot \hat{\gamma}_{k+i} \right) \\
m \left( \frac{\hat{v}_{y,k+i} - \hat{v}_{y,k+i-1}}{\Delta t} + \hat{v}_{x,k+i} \cdot \hat{\gamma}_{k+i} \right) \\
\end{bmatrix}
\end{equation}
\noindent 
where $\Delta t$ is the sampling interval, and $\hat{v}_{\cdot, k+i-1}$ denotes the state of the previous time step, which is initialized as the measured value $v_{\cdot, k}$ for the first step ($i=1$) and utilizes the recursively predicted state for subsequent steps ($i>1$).

On the other hand, $\boldsymbol{F}_{k+i}$ represents the resultant external forces acting on the vehicle body, calculated by aggregating the longitudinal and lateral forces of the tires projected onto the vehicle coordinate system. The total longitudinal and lateral tire forces are summations of the four individual wheels (e.g. $F_{xf} = F_{xfl} + F_{xfr}$).
\begin{equation}
\label{21}
\boldsymbol{F}_{k+i} = \begin{bmatrix}
(F_{xf,k+i}) \cos \delta_f - (F_{yf,k+i}) \sin \delta_f + F_{xr,k+i} \\
(F_{xf,k+i}) \sin \delta_f + (F_{yf,k+i}) \cos \delta_f + F_{yr,k+i}
\end{bmatrix}
\end{equation}
\noindent 
\textbf{Remark 2.} The tire forces are treated as known inputs. For production vehicles lacking direct force sensors, these values can be substituted with estimated tire forces without altering the structure of the proposed Koopman framework.

\item \textbf{Global Optimization Objective:} To synthesize the mathematical properties of the Koopman operator with the physical laws of vehicle motion, the network parameters are optimized through a multi-objective cost function. The total loss function $\mathcal{L}_{\text{Total}}$ and the parameter optimization problem are defined as follows:
\begin{equation}
\label{22}
\begin{split}
\mathcal{L}_{\text{Total}} &= \alpha_1 \mathcal{L}_{\text{Lin}} + \alpha_2 \mathcal{L}_{\text{Recon}} + \alpha_3 \mathcal{L}_{\text{Pred}} + \alpha_4 \mathcal{L}_{\text{DL}}
\end{split}
\end{equation}

\begin{equation}
\label{23}
\begin{bmatrix} \boldsymbol{\theta}_e & \boldsymbol{\theta}_d & \boldsymbol{A} & \boldsymbol{B} \end{bmatrix} =\arg \min_{\boldsymbol{\theta}_e, \boldsymbol{\theta}_d, \boldsymbol{A}, \boldsymbol{B}} \mathcal{L}_{\text{Total}}
\end{equation}
\noindent 
where $\alpha_{1,2,3,4}$ weight the contribution of each loss term.
\end{enumerate}

By embedding the force-equilibrium-based constraint ($\mathcal{L}_{\text{DL}}$) into the global optimization objective, the resulting model transcends purely statistical fitting to strictly adhere to the governing equations of vehicle motion. For brevity, this proposed Dynamic Loss-informed Deep Koopman framework is hereafter referred to as DLDK.

\begin{algorithm}[!t]
\caption{Offline Training of the Dynamic Loss-informed Deep Koopman (DLDK) Model}
\label{alg:offline_training}
\begin{algorithmic}[1]
\Require 
    Initialize $\boldsymbol{\theta}_e, \boldsymbol{\theta}_d, \boldsymbol{A}, \boldsymbol{B}$;
    $p, \alpha_{1, \dots, 6},Epoch_{max}$; Batch size $b_s$; Convergence threshold $\epsilon$.
\Ensure 
    Trained $\boldsymbol{\theta}_e, \boldsymbol{\theta}_d, \boldsymbol{A}, \boldsymbol{B}$.
\While{$Epoch < Epoch_{\max}$ \textbf{and} $\mathcal{L}_{\text{Val}} > \epsilon$}
    \For{each mini-batch of size $b_s$}
        \State Obtain lifted features $\boldsymbol\Phi(\boldsymbol{x}_k; \boldsymbol{\theta}_e)$.
        \State Calculate $\boldsymbol{z}_k$ with Eq.~\eqref{9}.
        \For{$i = 1$ to $p$}
            \State Linear evolution $\hat{\boldsymbol{z}}_{k+i}$ with Eq.~\eqref{18}.
            \State Recover state $\hat{\boldsymbol{x}}_{k+i}$ with Eq.~\eqref{18}.
            \State \parbox[t]{\dimexpr\linewidth-\algorithmicindent}{Compute $\hat{\boldsymbol{F}}_{k+i}$ with Eq.~\eqref{20}.}
            \State \parbox[t]{\dimexpr\linewidth-\algorithmicindent}{Accumulate $\mathcal{L}_{\text{Lin}}, \mathcal{L}_{\text{Recon}}, \mathcal{L}_{\text{Pred}}, \mathcal{L}_{\text{DL}}$ \\according to Eqs.~\eqref{15}--\eqref{17} and \eqref{19}.}
        \EndFor
        \State Compute $\mathcal{L}_{\text{Total}}$ via Eq.~\eqref{22}.
        \State \parbox[t]{\dimexpr\linewidth-\algorithmicindent}{Update $\boldsymbol{\theta}_e, \boldsymbol{\theta}_d, \boldsymbol{A}, \boldsymbol{B}$ via solving Eq.~\eqref{23} with \\ Adam optimizer \citep{kingma2014adam}.}
    \EndFor
    \State Evaluate $\mathcal{L}_{\text{Val}}$ on validation set.
    \State $Epoch \leftarrow Epoch + 1$
\EndWhile
\end{algorithmic}
\end{algorithm}

Since simultaneous optimization of the neural network parameters ($\boldsymbol{\theta}_e, \boldsymbol{\theta}_d$) and the system matrices ($\boldsymbol{A}, \boldsymbol{B}$) constitutes a non-convex problem, the traditional closed-form least squares solution is no longer applicable. Furthermore, minimizing the multi-step prediction loss requires backpropagating gradients through the recurrent structure of the rollout horizon. Therefore, we adopt the gradient-based Adam optimizer\citep{kingma2014adam} to solve Eq.~\eqref{23} iteratively. The detailed offline training procedure is summarized in Algorithm~\ref{alg:offline_training}.

\subsection{Online Adaptation: Theoretical Formulation and Algorithm}
\label{sec:online_theory}

Although the offline DLDK model captures the nominal dynamics, real-world operation introduces time-varying uncertainties. To enable robust adaptation, we strictly formulate the online learning problem and address the theoretical bottlenecks inherent to high-dimensional Koopman operators.

\subsubsection{Theoretical Analysis: Rank Deficiency and Ill-Posedness}
To analyze the stability of online adaptation, we first revisit the optimal offline solution established in Section~\ref{sec:2.2}, as formulated in Eq.~\eqref{13}. This formulation implicitly acknowledges that the covariance matrix may be singular. Let $\boldsymbol{\theta}_k = [\boldsymbol{z}_{k}^\top, \boldsymbol{u}_{k}^\top]^\top \in \mathbb{R}^d$ denote the instantaneous regressor vector at time step $k$. Consequently, the cumulative data matrix up to time $k$ is defined as $\boldsymbol{\Theta}_{1:k} = [\boldsymbol{\theta}_1, \dots, \boldsymbol{\theta}_k]$. The online recursive update relies on the inverse of the cumulative covariance matrix $\boldsymbol{R}_k$, defined as:
\begin{equation}
\label{24}
    \boldsymbol{R}_k = \boldsymbol{\Theta}_{1:k} \boldsymbol{\Theta}_{1:k}^\top
\end{equation}

However, strictly employing the inverse $\boldsymbol{R}_k^{-1}$ is structurally perilous. Although the dimension of the regressor is $d$, the system dynamics evolve on a low-dimensional manifold governed by the underlying physical degrees of freedom. Due to the nonlinear lifting of the neural network, $\boldsymbol{R}_k$ can span a space of dimension $r$ ($n+m < r \le d$). During practical driving scenarios, the effective numerical rank $r$ remains significantly lower than $d$ due to insufficient excitation. Consequently, $\boldsymbol{R}_k$ becomes inherently ill-conditioned or even rank-deficient.

Taking into account the singular value decomposition of $\boldsymbol{\Theta}_{1:k}$, the eigenvalues $\lambda_j$ of $\boldsymbol{R}_k$ exhibit a severe spectral gap after the $r$-th mode:
\begin{equation}
\label{25}
    \lambda_1 \ge \dots \ge \lambda_{r} \gg \lambda_{r+1} \approx \dots \approx \lambda_d \approx 0
\end{equation}

Consequently, the inverse matrix $\boldsymbol{P}_k = \boldsymbol{R}_k^{-1}$ becomes singular. By expanding $\boldsymbol{P}_k$ in the eigenvector $\mathbf{w}_j$:
\begin{equation}
\label{26}
    \boldsymbol{P}_k = \sum_{j=1}^{d} \frac{1}{\lambda_j} \mathbf{w}_j \mathbf{w}_j^\top = \underbrace{\sum_{j=1}^{r} \frac{1}{\lambda_j} \mathbf{w}_j \mathbf{w}_j^\top}_{\mathcal{S}: \text{Signal Subspace}} + \underbrace{\sum_{j=r+1}^{d} \frac{1}{\lambda_j} \mathbf{w}_j \mathbf{w}_j^\top}_{\mathcal{N}: \text{Null Space}}
\end{equation}

As $\lambda_{j} \to 0$ in the null space $\mathcal{N}$, the term $1/\lambda_j \to \infty$. This theoretically proves that matrix-inversion-based methods are ill-posed for Deep Koopman models, as any noise projected onto $\mathcal{N}$ causes unbounded parameter divergence. Therefore, in practical applications, traditional RLS-type algorithms need to be regularized by adding a small positive definite matrix (e.g., $\boldsymbol{R}_k \gets \boldsymbol{R}_k + \epsilon I$) to suppress the divergence, which requires careful tuning of $\epsilon$.

\subsubsection{Proposed Methodology: PI-VSS-NLMS}
To fundamentally avoid the numerical instability associated with ill-conditioned covariance matrices, we propose the Physics-Informed Variable Step-Size Normalized Least Mean Squares (PI-VSS-NLMS) algorithm, which circumvents matrix inversion entirely by leveraging the LMS framework.

\textbf{A. Recursive Problem Formulation}

First, we define the concatenated Koopman operator matrix at time step $k$ as $\boldsymbol{W}_k = [\boldsymbol{A}_k, \boldsymbol{B}_k] \in \mathbb{R}^{(d-m)\times d}$. The priori prediction error $\boldsymbol{e}_k$ is computed as:
\begin{equation}
\label{27}
    \boldsymbol{e}_k = \boldsymbol{z}_k - \boldsymbol{W}_{k-1}\boldsymbol{\theta}_{k-1}
\end{equation}

\textbf{B. Minimum-Norm Projection (Resolving Singularity)}

Instead of seeking a global inverse $\boldsymbol{P}_k$, which is structurally perilous when the effective numerical rank $r$ is far below the feature dimension $d$, we formulate the update as a sequence of constrained optimization problems based on the principle of minimal disturbance:
\begin{equation}
\label{28}
    \min_{\boldsymbol{W}_k} \|\boldsymbol{W}_k - \boldsymbol{W}_{k-1}\|_F^2 \quad \text{s.t.} \quad \boldsymbol{z}_k = \boldsymbol{W}_k \boldsymbol{\theta}_{k-1}
\end{equation}

By sequentially projecting the operator onto the hyperplanes defined by  $\boldsymbol{\theta}_{k-1}$, the algorithm achieves incremental adaptation while preserving history knowledge. Solving Eq.~\eqref{28} via Lagrange multipliers yields the NLMS projection update:
\begin{equation}
\label{29}
    \boldsymbol{W}_{proj} = \boldsymbol{W}_{k-1} + \mu \frac{\boldsymbol{e}_k \boldsymbol{\theta}_{k-1}^\top}{\|\boldsymbol{\theta}_{k-1}\|_2^2 + \epsilon}
\end{equation}
\noindent
where $\mu$ is the fixed step size and $\epsilon$ is a regularization constant to prevent division by zero.

A critical advantage of this formulation lies in its behavior under rank deficiency. As proven in \cite{sayed2003fundamentals}, for a rank-deficient linear system, the NLMS algorithm converges to the minimum-norm solution. This effectively approximates the operation of the pseudo-inverse without explicit matrix inversion. Consequently, the update naturally occurs within the signal subspace while suppressing the excitation of singular null space modes, providing implicit regularization against the parameter divergence that plagues RLS-type methods.

\textbf{C. Physics-Anchored Regularization (Suppressing Null-Space Drift)}

While NLMS prevents explosion, the singular null space remains unconstrained. In the absence of excitation (e.g., prolonged straight-line driving), even minute numerical residuals or measurement noises can induce a ``random walk'' of parameters in these unobserved dimensions. To ensure long-term numerical stability and physical consistency, we introduce a Physics Anchor mechanism, formulated as Tikhonov Regularization centered at the offline physical prior $\boldsymbol{W}_0$:
\begin{equation}
\label{30}
    \mathcal{J}_{reg} = \underbrace{\|\boldsymbol{z}_k - \boldsymbol{W}_k \boldsymbol{\theta}_{k-1}\|_2^2}_{\text{Data Fidelity}} + \rho_{anchor} \underbrace{\|\boldsymbol{W}_k - \boldsymbol{W}_0\|_F^2}_{\text{Physics Constraint}}
\end{equation}

The gradient descent step for the regularization term yields the physics-informed restoring force:
\begin{equation}
\label{31}
    \Delta \boldsymbol{W}_{PI} = -\rho_{anchor} (\boldsymbol{W}_{k-1} - \boldsymbol{W}_0)
\end{equation}

This formulation establishes a dynamic competition between two gradients. In the signal subspace $\mathcal{S}$ where excitation is persistent, the data-driven gradient dominates the update to capture real-time variations. Conversely, in the null space $\mathcal{N}$ where the data-driven gradient vanishes, the anchoring term acts as a steady restoring force that pulls the operator back toward the physically valid manifold $\boldsymbol{W}_0$. This dual-mechanism ensures that the adaptive operator remains bounded and physically interpretable throughout diverse driving cycles.

\textbf{D. Variable Step-Size (Noise Gating)}

To address the trade-off between the convergence rate and the steady-state misalignment, we construct a time-varying step size $\mu_k$ governed by an exponential modulation law based on the instantaneous error energy.
\begin{equation}
\label{32}
    \mu_k = \mu_{\min} + (\mu_{\max} - \mu_{\min}) \left[ 1 - \exp\left( -\frac{\eta_{\text{vss}}}{d-m} \|e_k\|_2^2 \right) \right]
\end{equation}
\noindent
where $\mu_{\max}$ and $\mu_{\min}$ define the dynamic range, and $\eta_{\text{vss}}$ is a sensitivity parameter. Mathematically, this acts as a soft noise gate. This mechanism ensures rapid tracking ($\mu_k \to \mu_{\max}$) during sudden dynamic changes and high-precision fine-tuning ($\mu_k \to \mu_{\min}$) during steady-state driving.

\textbf{Summary of Update Law:}
Combining these mechanisms, the final PI-VSS-NLMS update law is:
\begin{equation}
\label{33}
    \boldsymbol{W}_k = \boldsymbol{W}_{k-1} + \underbrace{\mu_k \frac{\boldsymbol{e}_k \boldsymbol{\theta}_{k-1}^\top}{\|\boldsymbol{\theta}_{k-1}\|_2^2 + \epsilon}}_{\text{Data-Driven Projection}} - \underbrace{\rho_{anchor} (\boldsymbol{W}_{k-1} - \boldsymbol{W}_0)}_{\text{Physics-Informed Anchor}}
\end{equation}

This formulation guarantees bounded stability by circumventing the numerical singularities associated with rank deficiency (via minimum-norm projection) and suppressing parameter drift within the singular null space (via physics-anchored regularization). The detailed procedure is presented in Algorithm~\ref{alg:online_adaptation}.

\begin{algorithm}[!t]
\caption{Online Adaptation via PI-VSS-NLMS}
\label{alg:online_adaptation}
\begin{algorithmic}[1]
\Require 
    $\boldsymbol{\theta}_e, \boldsymbol{W}_0, \mu_{\min}, \mu_{\max}, \eta_{\text{vss}}, \rho_{\text{anchor}}, \epsilon$.
\Ensure 
    Adapted $\boldsymbol{W}_k$.
\For{$k$ = 1: end}
    \State Obtain lifted features $\boldsymbol\Phi(\boldsymbol{x}_k; \boldsymbol{\theta}_e)$.
    \State Calculate $\boldsymbol{z}_k$ with Eq.~\eqref{9}.
    \State Calculate regressor vector $\boldsymbol{\theta}_{k-1}$
    \State Compute error $\boldsymbol{e}_k$ with Eq.~\eqref{27}.
    \State Update step size $\mu_k$ with Eq.~\eqref{32}.
    \State Update $\boldsymbol{W}_k$ with Eq.~\eqref{33}.
\EndFor
\end{algorithmic}
\end{algorithm}

\begin{figure}
	\centering
	\includegraphics{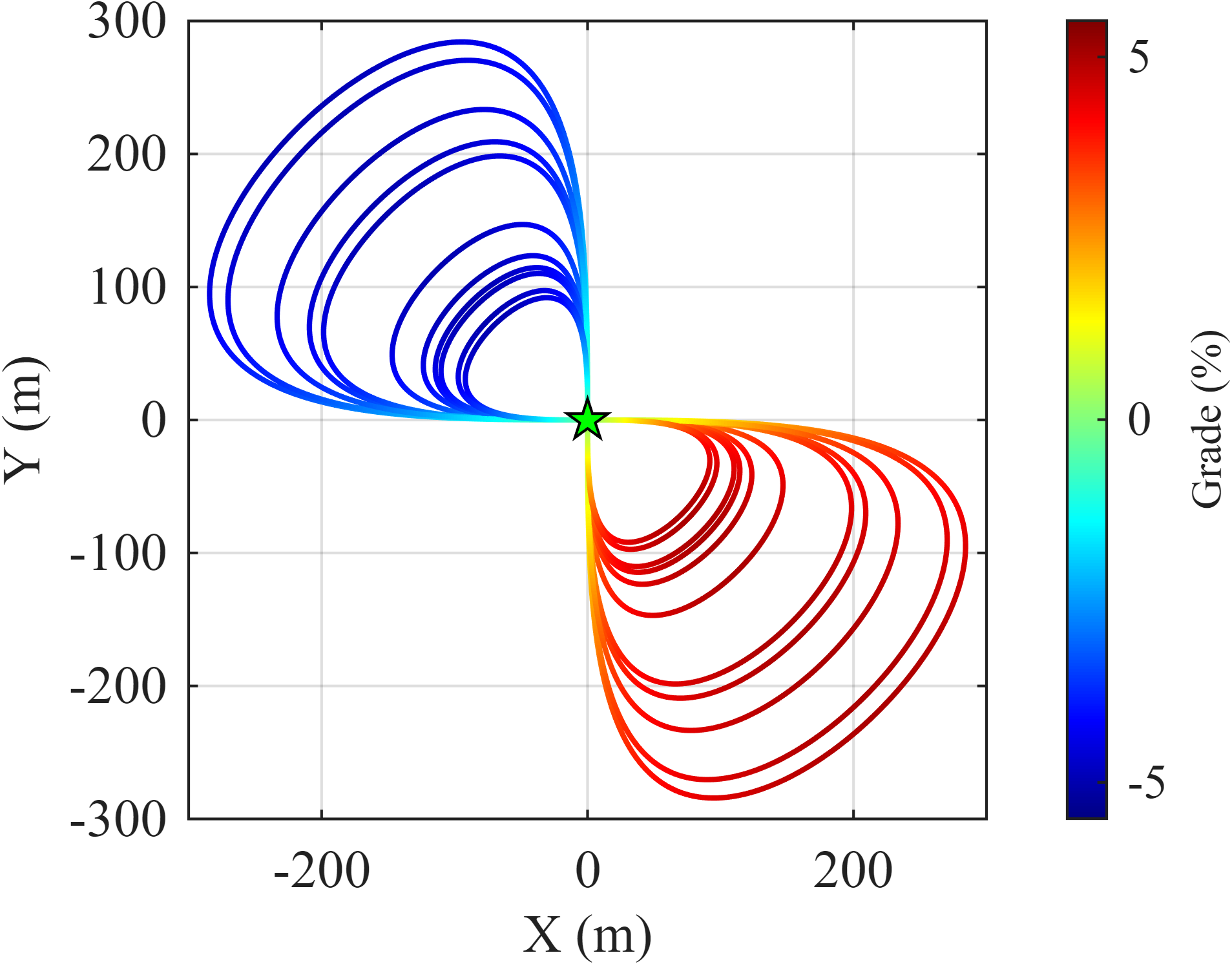}
	\caption{Procedurally generated stochastic training environment.}
	\label{FIG:3}
\end{figure}

\section{Validation and Discussion}
\label{sec:4}
This section evaluates the proposed framework in two stages: (1) offline benchmark tests against physics-based and data-driven baselines to verify modeling fidelity, (2) online validation under varied conditions to demonstrate the adaptability and robustness of the PI-VSS-NLMS algorithm, and (3) real-time Hardware-in-the-Loop (HIL) experiments on a dSPACE MicroAutobox III to confirm the computational efficiency and practical feasibility for embedded deployment.

\subsection{Data Generation and Preprocessing}
To ensure the Koopman model captures the full dynamic envelope, a stochastic procedural generation strategy is employed rather than standard maneuvers. As shown in Fig.~\ref{FIG:3}, the training dataset is derived from randomized 3D lemniscate (Figure-8) trajectories on a high-friction road ($\mu=0.85$). This setup provides two-fold excitation: (1) Lateral Dynamics: Randomly varying loop radii force continuous transitions between zero and peak yaw rates. (2) Vertical Coupling: A cosine elevation profile with a 5\% maximum grade introduces longitudinal load transfer, enriching the dataset with pitch-related tire force variations.

The generated coordinates were imported into CarSim 2024 for a 750 s simulation. With a 100 Hz sampling rate, 75,001 snapshots were collected and partitioned into training (70\%), validation (15\%), and testing (15\%) sets. Each snapshot consists of the state $\boldsymbol{x}_k = [v_x, v_y, \gamma]^\top$, the control input $\boldsymbol{u}_k = [T, \delta_{sw}]^\top$, and tire forces $\boldsymbol{F}_k$ required for the dynamic loss formulation. All features were standardized using min-max normalization \citep{islam2022application} to ensure training stability.

\begin{table}[width=\linewidth,cols=4,pos=t]
\caption{Key parameters of the D-class electric sedan in CarSim.}\label{tab:vehicle_params}
\begin{tabular*}{\tblwidth}{@{} LCCC@{} }
\toprule
\textbf{Parameter} & \textbf{Symbol} & \textbf{Value} & \textbf{Unit} \\
\midrule
Vehicle Mass & $m$ & 1525 & kg \\
Distance from c.g. to Front Axle & $l_f$ & 1.110 & m \\
Distance from c.g. to Rear Axle & $l_r$ & 1.756 & m \\
Height of center of mass & $h_g$ & 0.52 & m \\
Track Width & $w_B$ & 1.550 & m \\
Tire Effective Radius & $R_{eff}$ & 0.325 & m \\
Yaw Inertia & $I_z$ & 2315.3 & $\text{kg}\cdot\text{m}^2$ \\
Wheel Inertia & $I_w$ & 1.5 & $\text{kg}\cdot\text{m}^2$ \\
\bottomrule
\end{tabular*}
\end{table}

The target vehicle is a D-class electric sedan, with key vehicle parameters and training hyperparameters detailed in Table~\ref{tab:vehicle_params} and Table~\ref{tab:hyperparameters}, respectively. The co-simulation environment was implemented using CarSim 2024 and MATLAB R2025a, executed on a computing platform equipped with an AMD Ryzen 9 9955HX CPU and an NVIDIA RTX 5060 GPU.

\begin{table}[width=\linewidth,cols=2,pos=t]
\caption{Hyperparameters for Deep Koopman training and online adaptation.}\label{tab:hyperparameters}
\begin{tabular*}{\tblwidth}{@{} LC@{} }
\toprule
\textbf{Parameter} & \textbf{Value} \\
\midrule
\multicolumn{2}{l}{\textit{\textbf{Offline Training (Deep Koopman)}}} \\ 
Learning Rate & $ 10^{-4}$ \\
Batch Size & 1024 \\
Prediction Horizon $p$ & 50 \\
Discount Factor $\xi$ & 0.99\\
$\alpha_{1}$ & 0.1 \\
$\alpha_{2}$ & 0.1 \\
$\alpha_{3}$ & 1 \\
$\alpha_{4}$ & $10^{-4}$ \\
Auto-encoder & [$3 \quad 128\quad 128\quad 128\quad 16$] \\
Auto-decoder & [$16 \quad 128\quad 128\quad 128\quad 3$] \\
Matrix $\boldsymbol{A}$ Size & [$19 \quad 19$] \\
Matrix $\boldsymbol{B}$ Size & [$19 \quad 2$] \\
\midrule
\multicolumn{2}{l}{\textit{\textbf{Online Adaptation (PI-VSS-NLMS)}}} \\ 
Max Step Size $\mu_{\max}$ & 1.5 \\
Min Step Size $\mu_{\min}$ & 0.05 \\
Sensitivity Parameter $\eta_{\text{vss}}$ & 200 \\
Regularization constant $\epsilon$ & $1 \times 10^{-8}$ \\
Anchoring coefficient $\rho_{\text{anchor}}$ & $3 \times 10^{-3}$ \\
\bottomrule
\end{tabular*}
\end{table}

\subsection{Offline Modeling Accuracy Analysis}
\label{sec:4.2}

\begin{figure}
	\centering
	\includegraphics{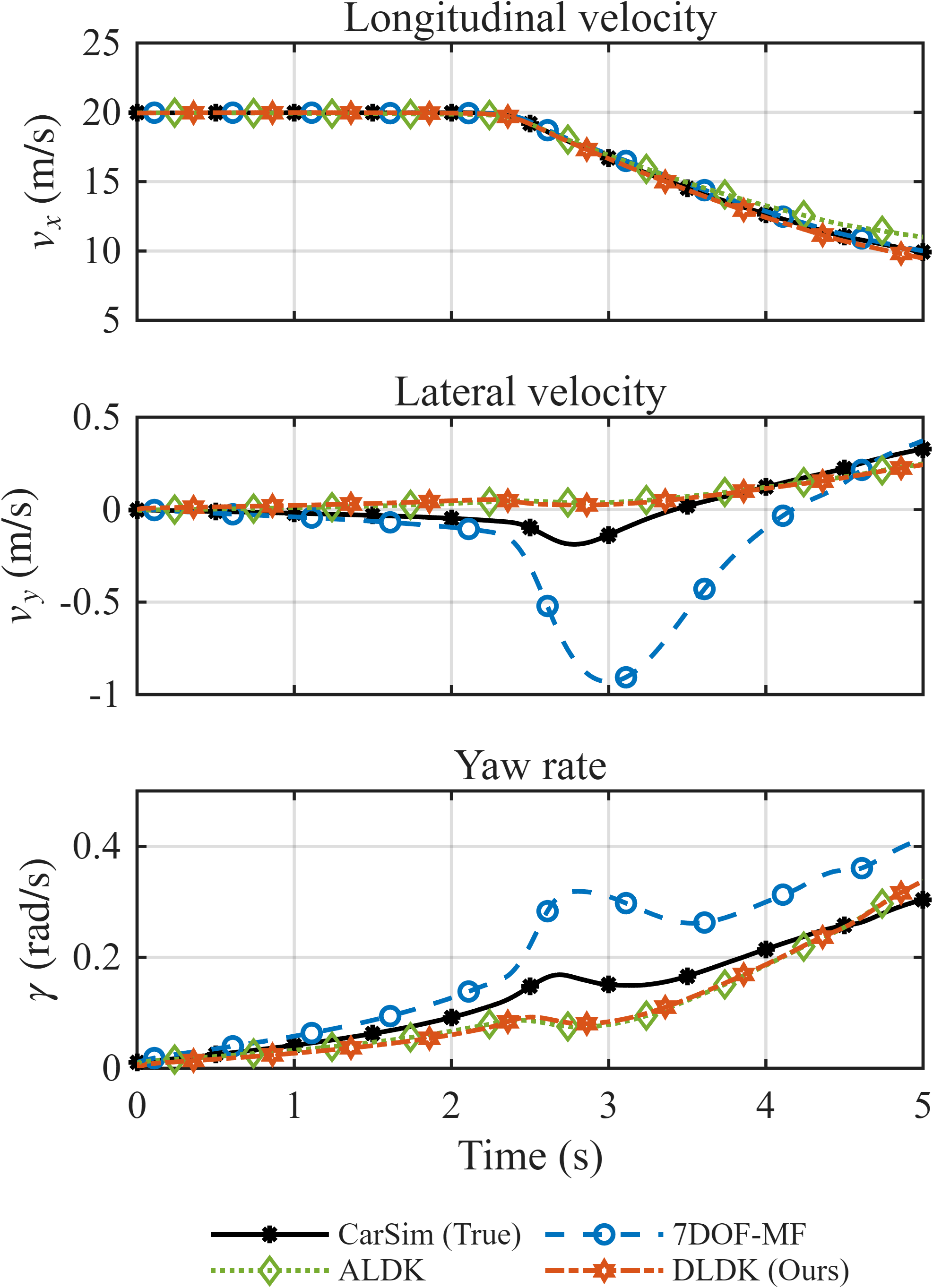}
	\caption{Open-loop prediction results over a 5-second horizon under extreme steering maneuvers.}
	\label{FIG:4}
\end{figure}

To evaluate short-term fidelity under extreme conditions, we extracted a 5-second segment involving a rapid steering maneuver from the test dataset. Fig.~\ref{FIG:4} visualizes the open-loop prediction results of the proposed DLDK compared against the physical benchmark (7DOF-MF) and the state-of-the-art kinematic competitor (ALDK).

\begin{table}[width=\linewidth,cols=5,pos=t]
\caption{Comparison of offline prediction accuracy (horizon = 5s).}\label{tab:offline_accuracy5s}
\begin{tabular*}{\tblwidth}{@{} LLCCC@{} }
\toprule
\multirow{2}{*}{\textbf{Model}} & \multirow{2}{*}{\textbf{Metric}} & $\boldsymbol{v_x}$ & $\boldsymbol{v_y}$ & $\boldsymbol{\gamma}$ \\
 & & (m/s) & (m/s) & (rad/s) \\
\midrule
\multirow{2}{*}{7DOF-MF} & RMSE & \textbf{0.1596} & 0.3271 & 0.0819 \\
 & Max & \textbf{0.3463} & 0.8106 & 0.1611 \\
\midrule
\multirow{2}{*}{ALDK} & RMSE & 0.4289 & 0.0872 & 0.0376 \\
 & Max & 1.0881 & 0.2246 & 0.0921 \\
\midrule
\multirow{2}{*}{\textbf{DLDK}} & RMSE & 0.1698 & \textbf{0.0868} & \textbf{0.0363} \\
 & Max & 0.4301 & \textbf{0.2125} & \textbf{0.0848} \\
\bottomrule
\end{tabular*}
\end{table}

Fig.~\ref{FIG:4} presents the open-loop prediction results, highlighting the distinct behaviors of the three benchmarks. The physics-based 7DOF-MF model suffers from a prominent phantom sideslip phenomenon; specifically, it predicts a transient $v_y$ drop to $-1.0$ m/s ($t$=3 s) while the ground truth remains stable at $-0.12$ m/s. While both data-driven models eliminate these non-physical artifacts, the ALDK exhibits a cumulative longitudinal drift, indicating that acceleration-based loss fails to fully account for cornering-induced resistance.

In contrast, the proposed DLDK maintains the highest fidelity across all states. By incorporating explicit tire-force constraints, the DLDK successfully captures cross-axis coupling, achieving a reduction of 60.4\% in $v_x$ RMSE compared to the ALDK. Crucially, the DLDK also significantly outperforms the 7DOF-MF in the lateral domain, reducing the $v_y$ and $\gamma$ RMSE by 73.5\% and 55.7\%, respectively. As summarized in Table~\ref{tab:offline_accuracy5s}, the DLDK constrains the maximum $v_y$ error to $0.2125$ m/s, avoiding the $0.8106$ m/s transient divergence seen in the 7DOF-MF and ensuring the error boundedness essential for safety-critical applications.

\subsection{Online Adaptation Performance}
\label{sec:4.3}

\begin{figure}
	\centering
	\includegraphics{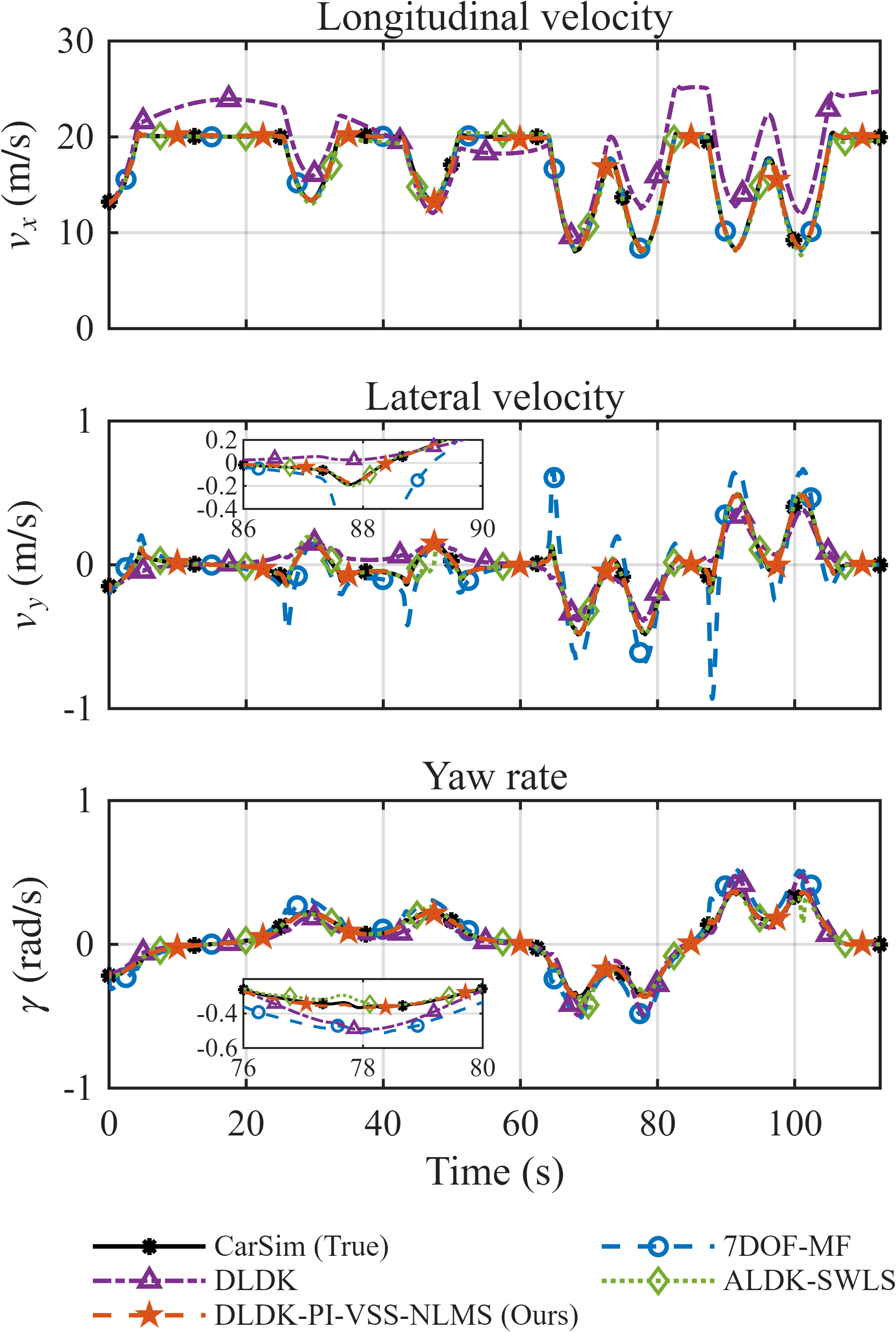}
	\caption{Online prediction results over the entire test cycle.}
	\label{FIG:5}
\end{figure}

Fig.~\ref{FIG:5} and Table~\ref{tab:online_performance} evaluate the adaptability and real-time feasibility of the proposed framework. Visually, the longitudinal velocity panel highlights the critical role of physics-anchored adaptation. The non-adaptive DLDK suffers from unbounded drift due to error accumulation. In contrast, the proposed DLDK-PI-VSS-NLMS strictly adheres to the ground truth, confirming that the update law successfully rectifies model mismatch and constrains predictions to the physical manifold. For lateral dynamics ($v_y$ and $\gamma$), while the ALDK-SWLS tracks well, the proposed method achieves superior transient handling, maintaining tighter alignment near local extrema (e.g., $t \approx 78$ s).

Quantitatively, the proposed method significantly outperforms the baselines. It reduces the lateral velocity RMSE by 76.7\% (from $0.0253$ m/s to $0.0059$ m/s) compared to the ALDK-SWLS and suppresses the longitudinal drift of the raw DLDK to a level comparable to the 7DOF-MF benchmark (RMSE: $0.1182$ m/s).

In addition to prediction accuracy, the proposed framework offers decisive computational advantages. While the RLS-based ALDK-SWLS requires $3.27$ ms/step due to the $\mathcal{O}(d^2)$ complexity of inverse covariance updates—which are also prone to numerical instability—the proposed algorithm relies solely on $\mathcal{O}(d)$ vector inner-products (Eq.~\eqref{27}). This yields a 5.6-fold speedup ($0.58$ ms/step) and eliminates singularity risks, guaranteeing both numerical robustness and high-frequency real-time feasibility for embedded vehicle control units.

\begin{table}[width=\linewidth,cols=5,pos=t]
\caption{Comparison of online prediction accuracy and computational efficiency (entire test cycle).}
\label{tab:online_performance}
\begin{tabular*}{\tblwidth}{@{} LLCCCC@{} }
\toprule
\multirow{2}{*}{\textbf{Model}} & \multirow{2}{*}{\textbf{Metric}} & $\boldsymbol{v_x}$ & $\boldsymbol{v_y}$ & $\boldsymbol{\gamma}$ & \textbf{Comp. Time} \\
 & & (m/s) & (m/s) & (rad/s) & (ms/step) \\
\midrule
\multirow{2}{*}{7DOF-MF} & RMSE & \textbf{0.0983} & 0.1216 & 0.0632 & \multirow{2}{*}{N/A}\\
 & Max & 0.3724 & 0.8106 & 0.1611 & \\
\midrule
\multirow{2}{*}{DLDK} & RMSE & 3.1847 & 0.0749 & 0.0428 & \multirow{2}{*}{--} \\
 & Max & 5.2230 & 0.2154 & 0.1411 & \\
\midrule
\multirow{2}{*}{ALDK-SWLS} & RMSE & 0.3252 & 0.0253 & 0.0198 & \multirow{2}{*}{3.27} \\
 & Max & 0.8095 & 0.1795 & 0.1998 & \\
\midrule
\multirow{2}{*}{\textbf{\makecell[l]{DLDK-PI-\\VSS-NLMS}}} & RMSE & 0.1182 & \textbf{0.0059} & \textbf{0.0038} & \multirow{2}{*}{\textbf{0.58}} \\
 & Max & \textbf{0.2395} & \textbf{0.0308} & \textbf{0.0164} & \\
\bottomrule
\end{tabular*}
\end{table}

\subsection{Robustness to Parameter Uncertainty}
\label{sec:4.4}
To evaluate the robustness of the proposed model, this section simulates a full-load scenario based on typical loading patterns for D-class sedans. Two rear passengers (75 kg each) and trunk cargo (175 kg) are added at distances of 0.8 m and 1.5 m behind the center of gravity, respectively. The increase in yaw inertia calculated using the parallel axis theorem \citep{gillespie2021fundamentals}:
$$\Delta I_z = 150 \times 0.8^2 + 175 \times 1.5^2 = 490\,\text{kg·m}^2$$
yielding $m_{\text{full}} = 1850$ kg (+21\%) and $I_{z,\text{full}} = 2805$ kg·m² (+21\%). The nearly proportional increases in mass and inertia reflect the realistic spatial distribution of passenger/cargo loads.

\begin{figure}
	\centering
	\includegraphics[width=\columnwidth]{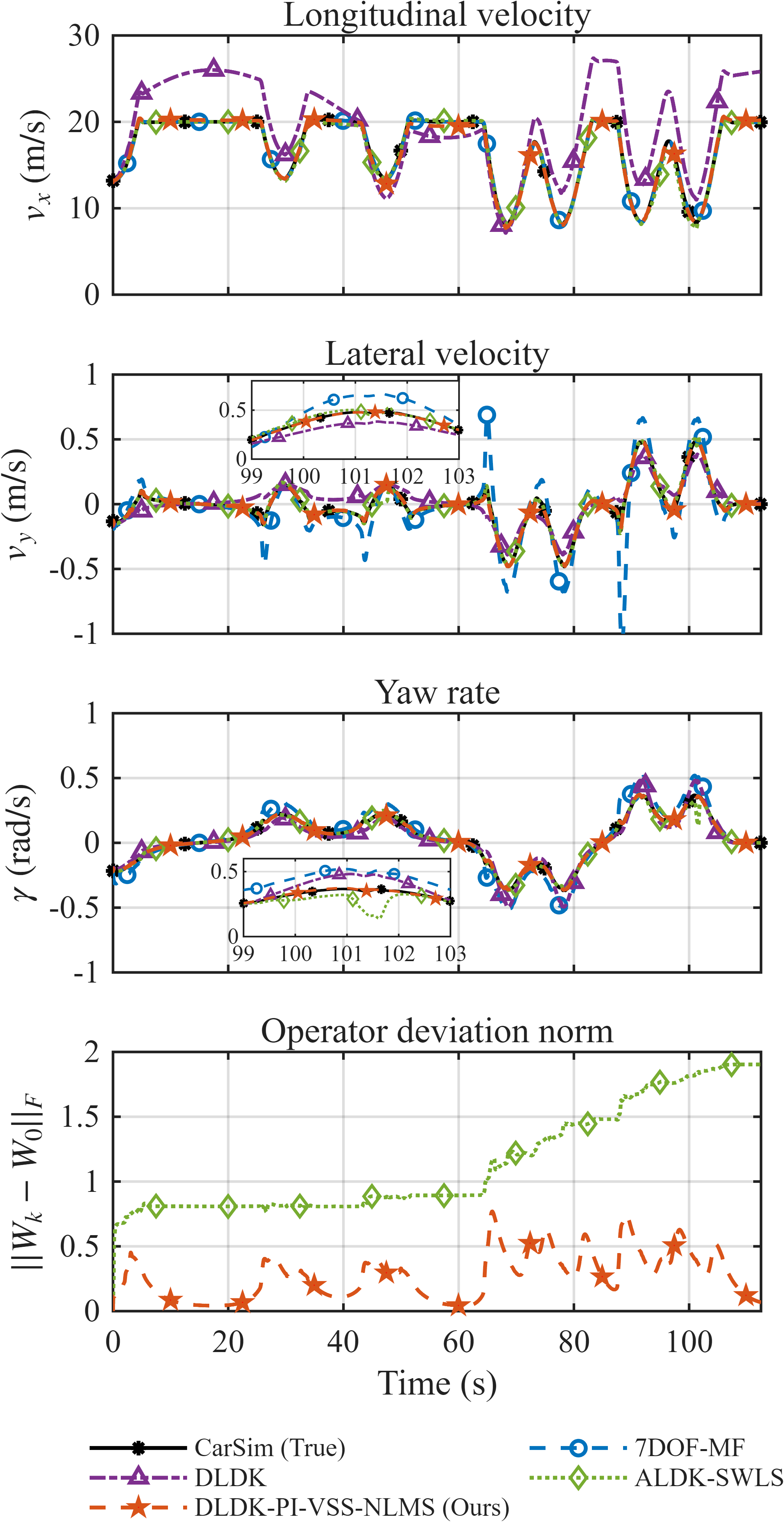}
	\caption{Robustness evaluation under full-load scenario ($m+21\%$, $I_z+21\%$).}
	\label{FIG:6}
\end{figure}

The tracking results in Fig.~\ref{FIG:6} highlight the limitations of non-adaptive methods. The physics-based 7DOF-MF exhibits massive deviations in $v_y$ and pronounced overshoot in $\gamma$, as fixed parameters do not balance the increased demands for momentum and lateral force. Meanwhile, the ALDK-SWLS suffers from significant phase lag and tracking errors during high-dynamic transitions ($t=99\text{--}103$ s) due to its windowed regression latency. In contrast, the proposed DLDK-PI-VSS-NLMS maintains precise synchronization with the ground truth across all states, demonstrating that the instantaneous gradient-based update law effectively compensates for parametric uncertainties in real-time.

The advantage of the physics-informed anchor is demonstrated in the operator deviation norm ($\|\boldsymbol{W}_k - \boldsymbol{W}_0\|_F$) panel. During low-excitation periods, such as the straight-line cruising segments (5-25 s and 50-65 s where $v_x$ is steady and $v_y \approx 0$), the deviation of the proposed method remains stable and close to zero. This confirms that the anchor provides the necessary restoring force, preventing disordered parameter updates typically caused by sensor noise in the absence of persistent excitation.

In contrast, as the maneuver intensity increases (specifically after 65 s), the deviation of the proposed method increases and fluctuates in response to the nonlinear dynamics under full-load conditions, but remains bounded. Meanwhile, the unconstrained ALDK-SWLS exhibits a continuous divergent drift with values approaching 2.0, indicating parameter instability within the rank-deficient null space. As quantified in Table~\ref{tab:robust_performance}, the balanced adaptation allows the proposed method to reduce the maximum error of the yaw rate by approximately 90\% compared to the sliding-window benchmark, ensuring high precision during aggressive maneuvers and robust stability during steady-state driving.

\begin{table}[width=\linewidth,cols=4,pos=t]
\caption{Quantitative performance comparison under the full-load scenario.}
\label{tab:robust_performance}
\begin{tabular*}{\tblwidth}{@{} LLCCC@{} }
\toprule
\multirow{2}{*}{\textbf{Model}} & \multirow{2}{*}{\textbf{Metric}} & $\boldsymbol{v_x}$ & $\boldsymbol{v_y}$ & $\boldsymbol{\gamma}$ \\
 & & (m/s) & (m/s) & (rad/s) \\
\midrule
\multirow{2}{*}{7DOF-MF} & RMSE & \textbf{0.0995} & 0.1326 & 0.0662 \\
 & Max & 0.3864 & 0.9661 & 0.1941 \\
\midrule
\multirow{2}{*}{DLDK} & RMSE & 4.0333 & 0.0779 & 0.0423 \\
 & Max & 7.1437 & 0.2359 & 0.1434 \\
\midrule
\multirow{2}{*}{ALDK-SWLS} & RMSE & 0.2672 & 0.0177 & 0.0158 \\
 & Max & 0.8883 & 0.1195 & 0.2255 \\
\midrule
\multirow{2}{*}{\textbf{\makecell[l]{DLDK-PI-\\VSS-NLMS}}} & RMSE & 0.2415 & \textbf{0.0077} & \textbf{0.0041} \\
 & Max & \textbf{0.5083} & \textbf{0.0452} & \textbf{0.0210} \\
\bottomrule
\end{tabular*}
\end{table}

\begin{figure}
	\centering
	\includegraphics{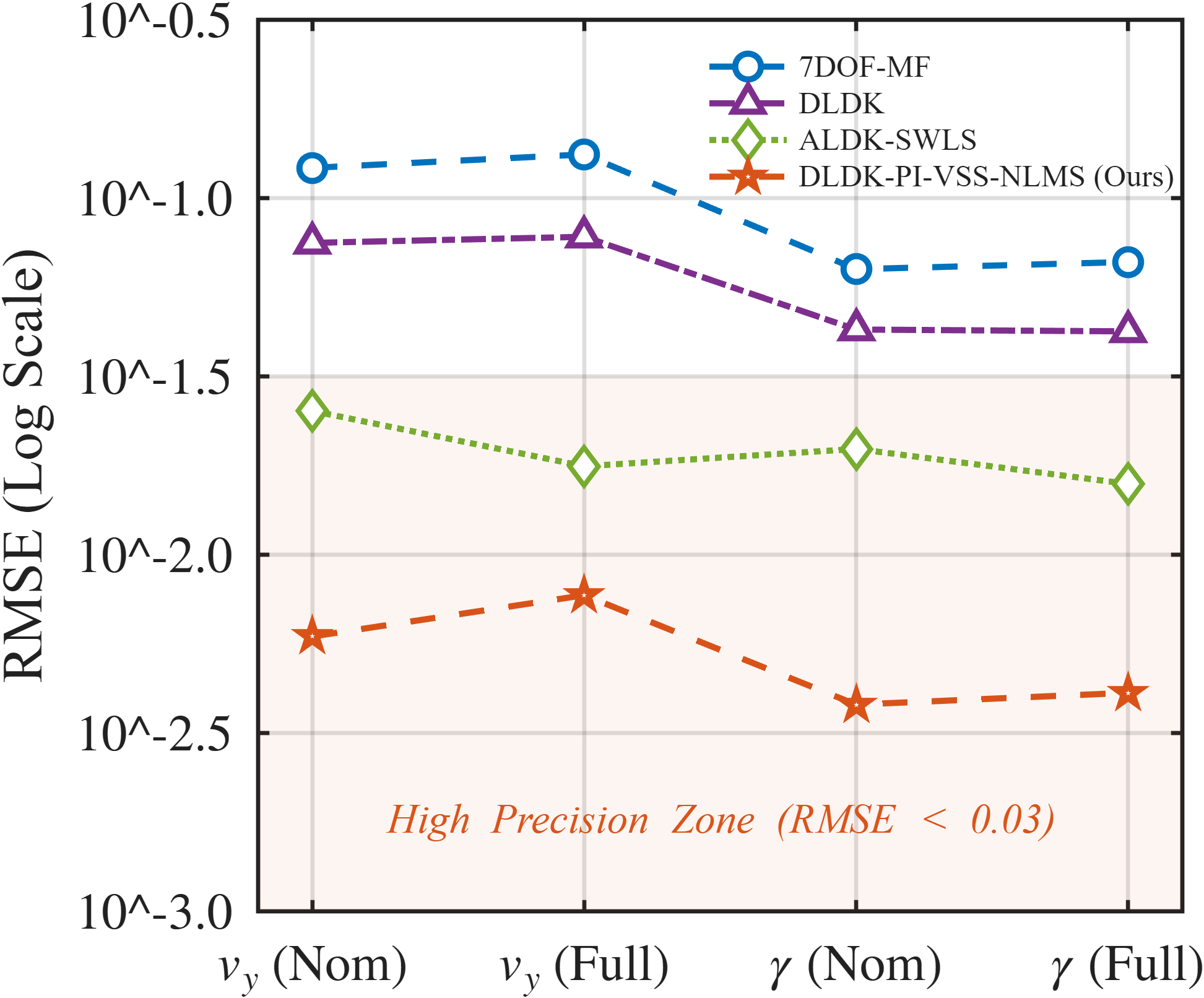}
	\caption{Logarithmic RMSE distribution of lateral dynamics under nominal and full-load conditions.}
	\label{FIG:7}
\end{figure}

To evaluate global adaptability, Fig.~\ref{FIG:7} presents the logarithmic RMSE distribution across varied scenarios. While the ALDK-SWLS hovers near the upper boundary of the High Precision Zone (RMSE $< 0.03$), the proposed DLDK-PI-VSS-NLMS anchors significantly deeper, maintaining a clear performance margin over the benchmarks. This confirms that the proposed framework maintains fidelity and robust consistency against parametric uncertainty.

\subsection{Generalization Test}
\label{sec:4.5}
To evaluate generalization beyond the unstructured Lemniscate training distribution, an ISO 7401 open-loop sinusoidal test (0.5 Hz at 80 km/h) is conducted. Unlike the quasi-random excitations in the training set, this structured periodic maneuver verifies whether the learned Koopman operator captures the intrinsic system eigenvalues or merely overfits irregular training patterns.

\begin{figure}
	\centering
	\includegraphics{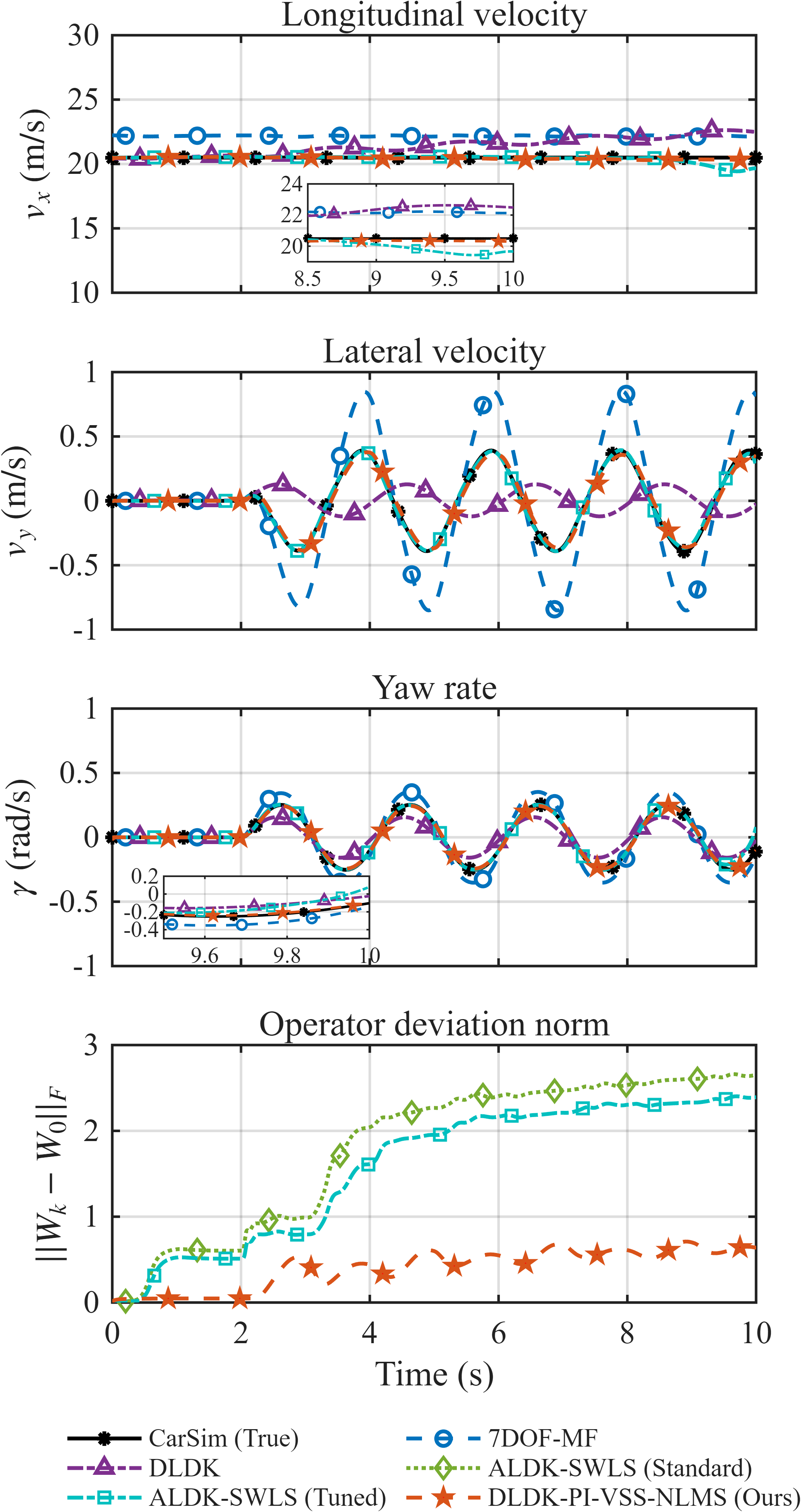}
	\caption{Generalization evaluation under standardized sinusoidal excitation (0.5 Hz).}
	\label{FIG:8}
\end{figure}

The generalization performance is visualized in Fig.~\ref{FIG:8}. Divergence occurs in the $v_x$ panel ($t=8.5\text{--}10$ s) and the $\gamma$ panel ($t=9.5\text{--}10$ s). While the proposed DLDK-PI-VSS-NLMS maintains robust synchronization with the ground truth, the ALDK-SWLS (Standard) suffers from catastrophic numerical divergence. Even with a doubled regularization parameter ($\epsilon = 1.66 \times 10^{-2}$), the ALDK-SWLS (Tuned) still exhibits visible drift. This failure indicates that sliding-window regression struggles with error accumulation under sustained high-frequency excitation; once the windowed data become ill-conditioned, the estimator loses convergence. In contrast, the proposed method remains stable, validating that the instantaneous gradient-based update law effectively filters numerical instability.

\begin{table}[width=\linewidth,cols=4,pos=!tb]
\caption{Quantitative performance comparison under the generalization test scenario.}
\label{tab:Generalization_performance}
\begin{tabular*}{\tblwidth}{@{} LLCCC@{} }
\toprule
\multirow{2}{*}{\textbf{Model}} & \multirow{2}{*}{\textbf{Metric}} & $\boldsymbol{v_x}$ & $\boldsymbol{v_y}$ & $\boldsymbol{\gamma}$ \\
 & & (m/s) & (m/s) & (rad/s) \\
\midrule
\multirow{2}{*}{7DOF-MF} & RMSE & 0.0995 & 0.1326 & 0.0662 \\
 & Max & 0.3864 & 0.9661 & 0.1941 \\
\midrule
\multirow{2}{*}{DLDK} & RMSE & 1.1424 & 0.2933 & 0.0703 \\
 & Max & 2.1327 & 0.4758 & 0.1118 \\
\midrule
\multirow{2}{*}{\textbf{\makecell[l]{ALDK-SWLS\\(Tuned)}}} & RMSE & 0.2719 & \textbf{0.0118} & 0.0212 \\
 & Max & 1.0651 & 0.0881 & 0.1821 \\
\midrule
\multirow{2}{*}{\textbf{\makecell[l]{DLDK-PI-\\VSS-NLMS}}} & RMSE & \textbf{0.0952} & 0.0325 & \textbf{0.0078} \\
 & Max & \textbf{0.1995} & \textbf{0.0497} & \textbf{0.0121} \\
\bottomrule
\end{tabular*}
\end{table}
The operator deviation norm in the bottom panel demonstrates how early-stage parameter stability influences late-stage tracking convergence. During the initial low-excitation phase ($0-2$ s), characterized by steady straight-line cruising, the proposed DLDK-PI-VSS-NLMS maintains a deviation near zero, confirming that the physics-informed anchor suppresses disordered updates in the absence of persistent excitation. In contrast, the ALDK-SWLS initiates disordered parameter updates immediately from $t=0$ s, driven by noise within the null space. The early-stage ``random walk'' of parameters accumulates a significant structural offset, which serves as a biased initial condition that accelerates numerical instability upon entering the high-dynamic sinusoidal phase ($2-10$ s). Consequently, the proposed method adaptively updates the Koopman operator within a bounded manifold. In contrast, the ALDK-SWLS suffers from cumulative parameter drift, which causes the divergence observed in the $v_x$ and $\gamma$ panels toward the end of the simulation. This comparison shows that by preventing unconstrained updates during low-excitation intervals, the proposed method ensures long-term robustness even under sustained high-frequency maneuvers.

\begin{figure}
	\centering
	\includegraphics{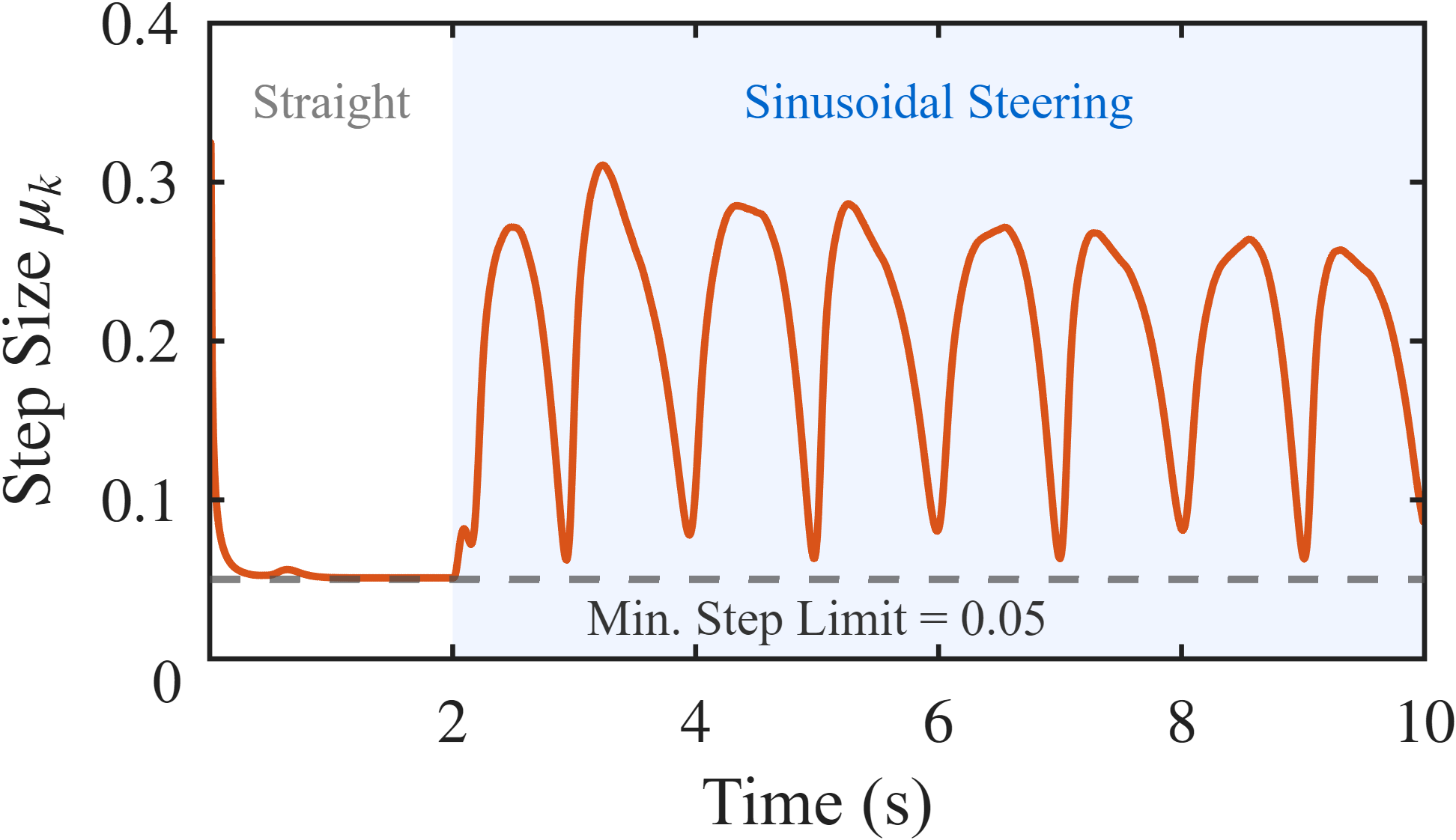}
	\caption{Evolution of the adaptive step size $\mu_k$ under sinusoidal excitation.}
	\label{FIG:9}
\end{figure}

Table~\ref{tab:Generalization_performance} further supports these findings. The proposed method achieves a $v_x$ RMSE of $0.0952$ m/s and reduces the maximum error of $\gamma$ to $0.0121$ rad/s-a tenfold reduction compared to the ALDK-SWLS (Tuned) ($0.1821$ rad/s). The improvement in precision correlates with the adaptive step size $\mu_k$ (Fig.~\ref{FIG:9}), which exhibits a clear periodic pattern at 0.5 Hz. By dynamically scaling $\mu_k$ to approximately $0.3$ during high-dynamic peaks and recovering to lower bound $0.05$ during quasi-straight phases, the framework achieves an balance between rapid error correction and noise suppression.

\subsection{Real-time Hardware-in-the-Loop Validation}
\begin{figure}
	\centering
	\includegraphics[width=\linewidth]{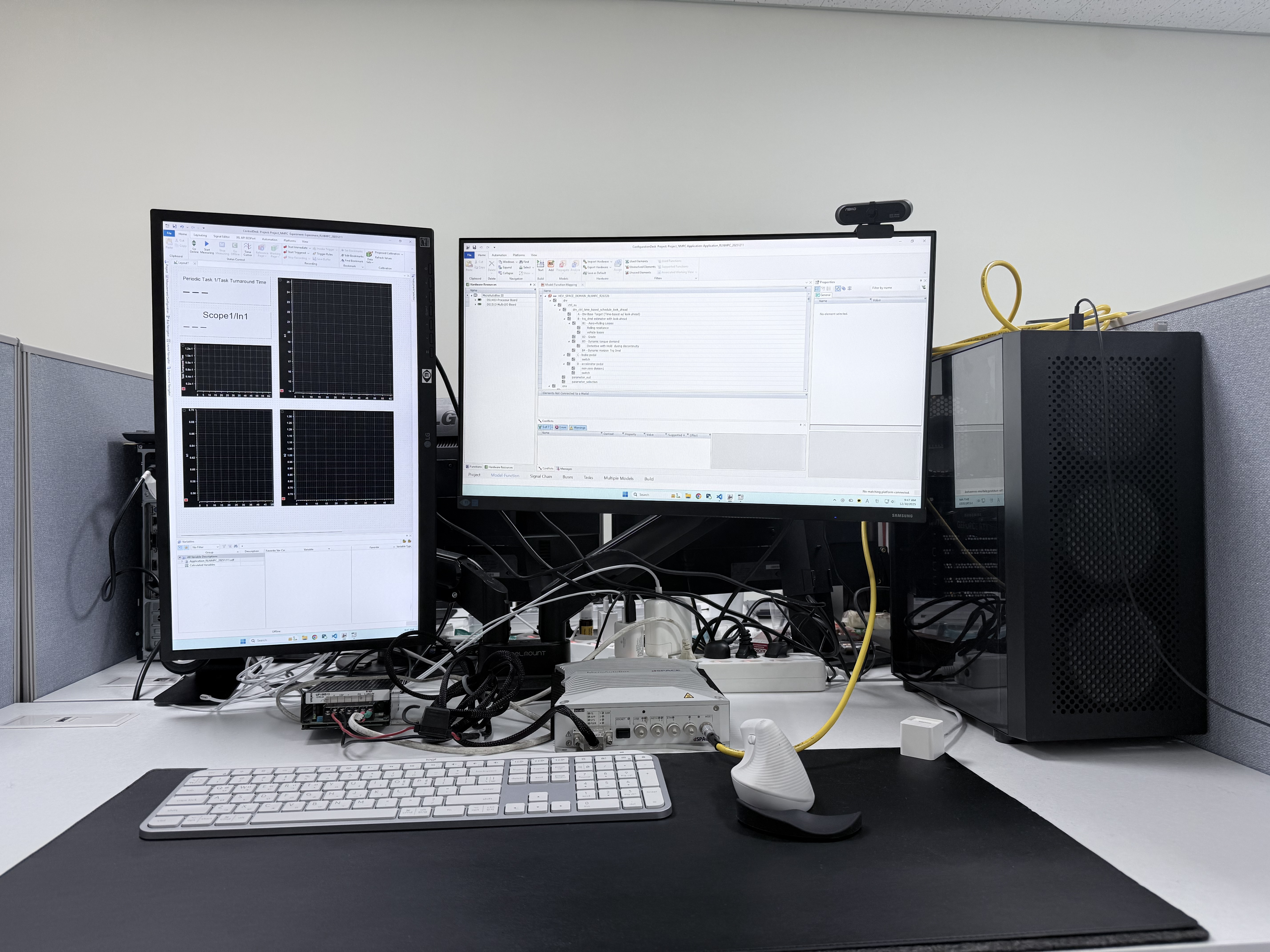}
	\caption{HIL experimental setup based on dSPACE MicroAutobox III.}
	\label{FIG:10}
\end{figure}

To further evaluate the practical feasibility of the proposed DLDK-PI-VSS-NLMS algorithm, a HIL experiment is conducted. The control algorithm is deployed onto a dSPACE MicroAutobox III real-time controller, as shown in Fig.~\ref{FIG:10}, representing a Rapid Control Prototyping setup.

\begin{figure}
	\centering
	\includegraphics{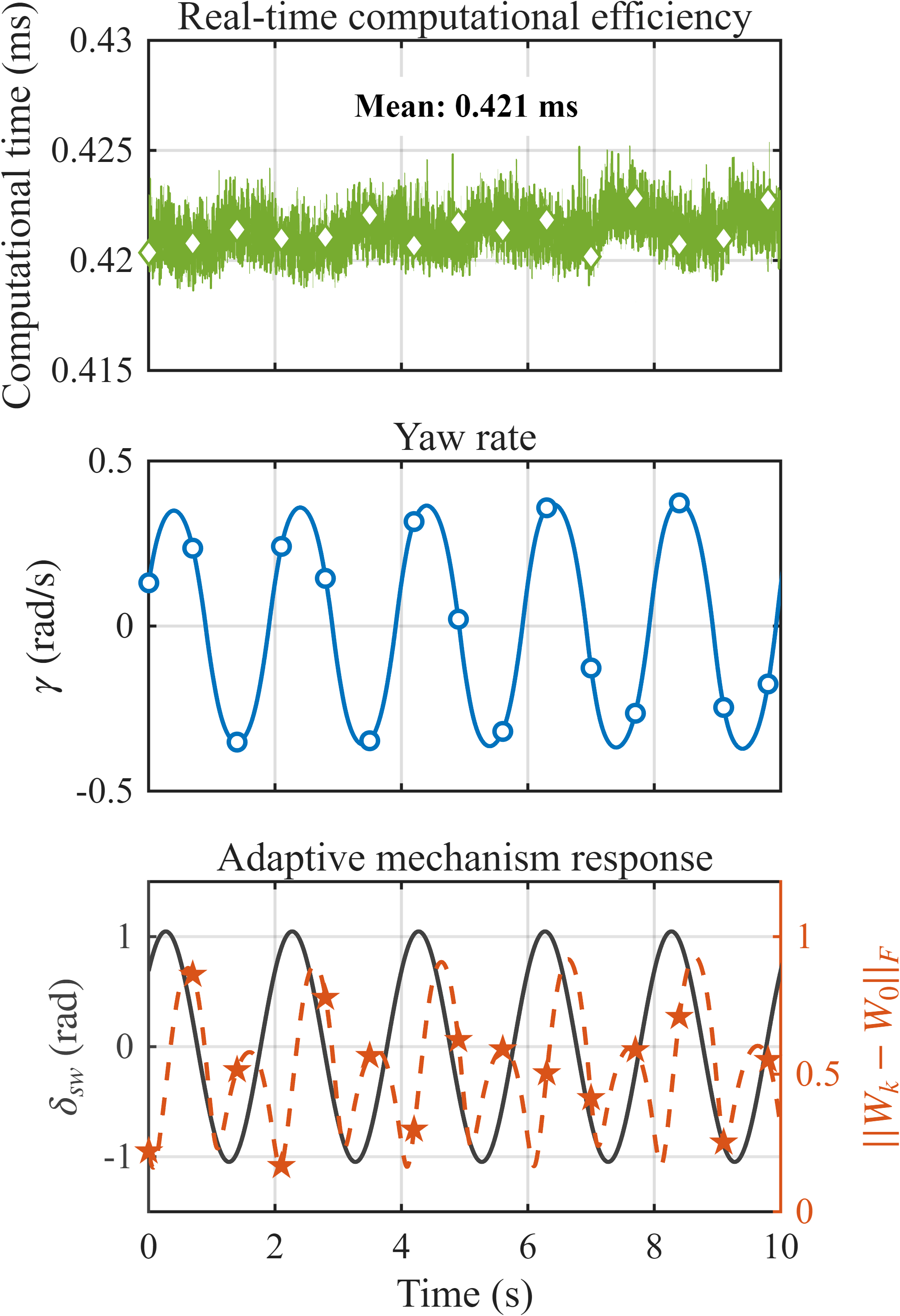}
	\caption{Real-time HIL validation results of the DLDK-PI-VSS-NLMS algorithm.}
	\label{FIG:11}
\end{figure}

The real-time validation results are illustrated in Fig.~\ref{FIG:11}, with the simulation environment configured with a fixed step size of 1 ms. The top panel presents the real-time execution time, revealing an average computational time of approximately 0.421 ms, which is well within the 1 ms sampling interval and confirms that the proposed model is computationally efficient enough for high-frequency vehicle dynamics control. Furthermore, the yaw rate ($\gamma$) and the steering input ($\delta_{sw}$) profiles indicate that the model remains stable and accurate under dynamic maneuvers. Finally, the bottom panel depicts the evolution of the adaptive mechanism, where the operator deviation norm ($\|\boldsymbol{W}_k - \boldsymbol{W}_0\|_F$) responds promptly to the varying vehicle states, ensuring the Koopman operator accurately captures the nonlinear tire-force-driven dynamics in real-time.

In summary, the HIL test validates that the proposed method can be successfully implemented on standard automotive grade hardware with a sufficient safety margin for real-time processing.

\section{Conclusion}
\label{sec:5}
This work develops and validates a physics-informed adaptive Deep Koopman framework to resolve the conflict between model fidelity and numerical stability in vehicle dynamics modeling under extreme conditions. Superior to purely kinematic-based learning approaches, the proposed tire-force-driven architecture embeds 7DOF dynamic equilibrium constraints, ensuring that the learned observables are physically interpretable rather than mathematical abstractions. To address the inherent rank deficiency in the lifted state space, a PI-VSS-NLMS algorithm is designed based on minimum-norm solution theory. This projection-based mechanism effectively eliminates the singularity accumulation and parameter drift observed in conventional RLS methods, guaranteeing operator stability. HIL validation on a dSPACE MicroAutobox III platform confirms the algorithm's real-time feasibility with an average execution time of 0.421 ms. The experimental results demonstrate that the proposed framework realizes high-precision trajectory tracking and robust adaptability.

In future work, we will attempt to integrate this real-time linearizable model into the MPC framework. The focus will be on leveraging the updated Koopman operator to enhance the stability control of autonomous vehicles  during limit handling scenarios.
\printcredits

\section*{Data availability}
Data will be made available on request.
\section*{Funding}
This work was supported by the National Natural Science Foundation of China under Grant 52372376; and the Graduate Research and Innovation Foundation of Chongqing under Grant CYB240012; and the China Scholarship Council under Grant 202506050050.
\section*{Declaration of competing interest}
The authors declare that they have no known competing financial interests or personal relationships that could have appeared to influence the work reported in this paper.

\bibliographystyle{cas-model2-names}
\bibliography{References}

\end{document}